\documentclass[12pt,preprint]{aastex}

\usepackage{times}

\shorttitle{Variation of the Solar Diameter}

\begin{document}

\title{Variation of the Diameter of the Sun
as Measured by the Solar Disk Sextant (SDS)}

\author{S. Sofia\altaffilmark{1}, 
        T.~M.~Girard\altaffilmark{1},
        U.~J.~Sofia\altaffilmark{2},
        L.~Twigg\altaffilmark{3},
        W.~Heaps\altaffilmark{4},
        G.~Thuillier\altaffilmark{5}
       }

\altaffiltext{1}{Astronomy Department, Yale University, P.O. Box 208101,
New Haven, CT 06520-8101, USA, 
sabatino.sofia@yale.edu, terry.girard@yale.edu}
\altaffiltext{2}{Physics Department, American University,
Washington, DC 20016, USA}
\altaffiltext{3}{Science Systems and Applications Inc., NASA/GSFC,
Greenbelt, MD, USA}
\altaffiltext{4}{Instrument Technology Center, NASA/GSFC,
Greenbelt, MD, USA}
\altaffiltext{5}{LATMOS-CNRS, 11 boulevard d'Alembert, 78280 Guyancourt,
France}

\begin{abstract}
The balloon-borne Solar Disk Sextant (SDS) experiment has measured the
angular size of the Sun on seven occasions spanning the years 1992
to 2011.
The solar half-diameter --
observed in a 100-nm wide passband centred at 615 nm --
is found to vary over that period by up to 200 mas,
while the typical estimated uncertainty of each measure is 20 mas.
The diameter variation is
{\it not} in phase with the solar activity cycle; thus, the
measured diameter variation
cannot be explained as an observational artefact of surface activity.
Other possible instrument-related explanations for the observed variation are 
considered but found unlikely, leading
us to conclude that the variation is real.
The SDS is described here in detail, as is the complete analysis procedure
necessary to calibrate the instrument and allow comparison of diameter measures
across decades.

\end{abstract}



\section{Introduction}

Measuring the solar diameter (and its variations) has a long and controversial
history with results highly inconsistent with each other, not only in absolute
values, but also in trends with time.
Detailed reviews of these efforts are presented in
Djafer et al.~(2008) and
Thuillier et al.~(2005, 2006)
and will not be repeated here.
More specifically, here we will only address direct measurements, as opposed to
values derived from the analysis of historical data (e.g. duration of solar
eclipses, timings of transits of Mercury, etc.).
A summary compilation of radius measurements from 1660 to the present
era (Pap et al.~2001) shows differences in excess of 2 arcsec, which
persist to recent simultaneous measurements.
See for instance Table 1 of Djafer et al.~(2008), which summarizes
such direct measurements, showing that they exhibit
differences of up to 0.8 arcsec while claiming internal accuracy of a few
tens of milli-arcseconds.
The main conclusion of the paper by Djafer et al. was that measuring the
solar diameter is a very difficult undertaking, and that the principal causes
of the inconsistent results
are the effects of the Earth's atmosphere, differences in
the definition of the limb edge, the spectral range of the measurements, and
instrumental differences that cannot be independently calibrated with a
precision higher than the expected variations.

On the more specific issue of variability, the results are even more extreme 
(Thuillier et al.~2005), varying from 1000 mas amplitude and in phase with solar 
activity (No\"{e}l 2004), to 200 mas and out of phase (Delmas \& Laclare 2002,
Egidi et al.~2006), to no significant change (Brown \& Christensen-Dalsgaard 1998, 
Wittmann 2003,
Penna et al.~2002, Kuhn et al.~2004).

Whereas differences of the results between the different diameter
determinations can arise from the various edge definition algorithms, and the
range of wavelength of measurements, the extreme differences in trends can only
arise from atmospheric effects, and/or from the large instrumental effects
produced by the extreme environment of Sun-pointing telescopes that do not
have a system for internal scale calibration.

This is best illustrated by No\"{e}l (2004) where it is shown that concurrent
measurements made with the same type of instrument (Danjon astrolabes) located
at Calern, (France), Santiago, (Chile), and Rio de Janeiro (Brazil) not
only show differences of up to $0.4''$, but also differ in trend with time.
The Calern data exhibit an anticorrelation with the solar activity cycle, while
the South American data show a positive correlation.

If we confine our discussion to space borne (or space-like) measurements, the
atmospheric effects are removed, or greatly reduced, and what remains are the
instrumental effects.
This selection leaves only 5 experiments for which diameter measurements
have been attempted:  the MDI on SOHO, the HMI on SDO, RHESSI, the SODISM
experiment on PICARD, and the SDS.
Of those, only the SDS and SODISM have internal calibration capability.
Because of the absence of such calibration, the RHESSI results only presented
determinations of the
solar oblateness, but not of the diameter
(Fivian et al.~2008).
This was based on the reasonable assumption that the instrumental scale would
not change significantly during the short interval in which an instrumental
rotation is performed.
Angular diameter determinations with SOHO/MDI
(Bush et al.~2010)
find no discernible variation, although
without an internal angular calibration these measurements must rely on
other means of correcting significant long-term changes in the instrument.
As with RHESSI, SDO/HMI has made observations sufficient to measure the
solar oblateness (Fivian et al.~2012), although results of these measures have
yet to appear.
SDO has not been operating long enough to address the issue of long-term solar
diameter variation.
PICARD/SODISM (Thuillier et al.~2011, Assus et al.~2008)
is still in a validation phase, and no results have been
published to date.
As a consequence, the only currently available results from instruments in a
space-like environment and with internal calibration, are the SDS results
presented here.

Because the SDS is balloon-borne, it can only be flown during periods when the
high-velocity stratospheric winds change direction (twice per year) at which
time, and for a few days, their velocity is low (the so called turnaround
period).
During turnaround we can have day-long flights and still remain within the
range of control of the NASA/Columbia Scientific Balloon Facility (CSBF), the
organization that operates these flights.
Moreover, at the location of the CSBF base in Fort Sumner, New Mexico,
turnaround occurs
in early May and late September.
At the spring turnaround, the Sun is too high in the sky, so that for several
hours around local noon the Sun would be behind the balloon, and hours of
observation would be lost.  As a consequence, flights are not recommended at
that time.
Hence, only fall flights are desirable.
More precisely, since the SDS in the current gondola cannot tilt above
53$^{\circ}$, those flights have to take place on September 25 or later.

Although we have flown the SDS eleven times, starting in the late 1980s, the
instrumental configuration required to reach a precision of tens of milliarc
sec (to be explained in Section~\ref{S-SI})
was only achieved starting with Flight~6,
in 1992.
As a consequence, this paper only presents the results of the seven flights
since and including Flight~6.

We describe the SDS instrument and its operation in
Sections~\ref{S-SI} and \ref{S-O}, respectively.
Section~\ref{S-DA}
details the data-analysis procedures and pipeline.
Section~\ref{S-R}
presents the diameter results and Section~\ref{S-D} discusses them, while
Section~\ref{S-C} provides a concluding summary.

\section{SDS Instrument}
\label{S-SI}

\begin{figure}
\centerline{\includegraphics[width=0.8\textwidth,clip=,]{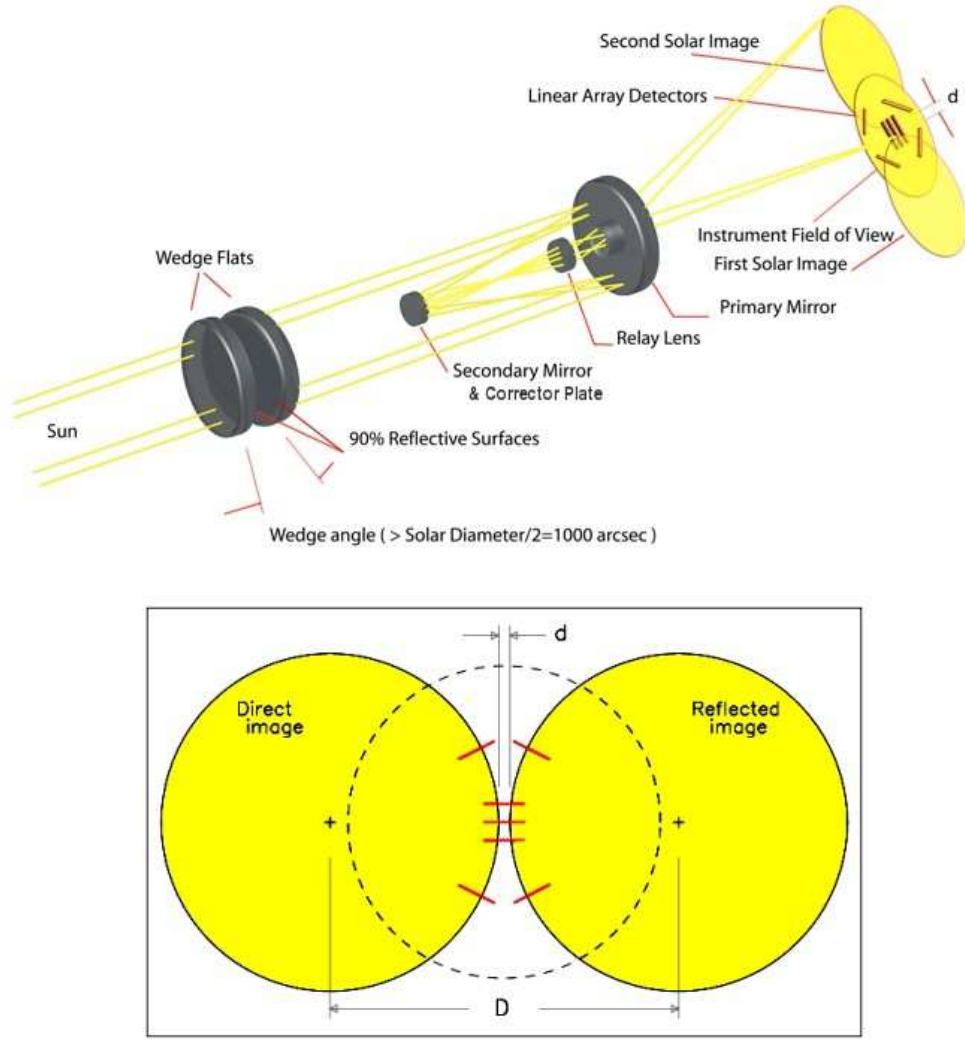}}
\caption{
{\it Upper:} Optical layout of the SDS.
The SDS is built around a ruggedized Questar telescope supplemented with
a beam splitting wedge and a Barlow relay lens to increase the focal length.
The beam splitting wedge produces a direct image and a reflected
image that is offset by slightly more than the Sun's diameter.
The solar limb is measured along ten radials by seven linear-array CCDs
mounted in the focal plane.
{\it Insert:} Detail of the detector plane.
The red line segments indicate the location of the CCDs; the
dotted circle shows the approximate field of view.
The solar diameter is determined by precise measurement of the disks'
separation, $D$, and gap, $d$.
}
\label{F-layout}
\end{figure}

The current version of the SDS balloon borne instrument is a package that has
been successfully flown several times on 12 - 29 Mcf (million cubic ft)
balloons.
The basic principle of the SDS instrument is the use of a mechanically and
optically stable beam splitting wedge (BSW) as an angle reference to form a
double image of the Sun separated by slightly more than its angular diameter.
The constancy of the angle of the BSW is achieved by utilizing molecular
contact fabrication techniques.
It can be shown that by measuring the small distance between images,
$d$, one can achieve the necessary accuracy much more easily than if one
attempts to measure the full diameter directly.
The level of dimensional stability required within the focal plane is
relaxed as $d/D$, where $D$ is the distance between the centre of two solar
images; in our case $d\sim6$ mm and $D\sim200$ mm.
Figure~\ref{F-layout} illustrates the measurement concept and the layout of
the detectors in the SDS focal plane.

This method also allows one to monitor and correct for instrument changes
(e.g. focal length changes, etc.) in a straightforward way.
Changes in any of the optical components downstream of the wedge will
affect both images and can be calibrated out.
The instrument consists of three basic items, the optical system, the
assembly of linear-array detectors, and the electronics package mounted
separately from the telescope.
The accuracy of the SDS derives from the system design, which uses a single
optical train to transfer the split solar images to the detectors.

In the current instrument, $F$ = 20.5 m, and $2W = 1978.94'' $. Further
details of the instrumental properties are given in
Sofia et al.~(1991) and summarized in Table~\ref{T-specs}.

\begin{table}
\caption{
SDS characteristics}
\label{T-specs}
\begin{tabular}{ll}
\hline
  focal length :  & 20.5 m  \\
  aperture :      & 12.7 cm  \\
  passband :      & $\lambda_o$ = 615 nm, $\delta\lambda \sim$ 100 nm \\
  pixel scale :   & 0.13 arcsec/pixel \\
  wedge angle $\times$2 :   & 1978.94 arcsec \\
\hline
\end{tabular}
\end{table}

\subsection{Optical system}
\label{S-SIa}

The stability of the optics needed by this instrument is of the level commonly
used in optical interferometry.
For this reason, the balloon flight version currently utilizes similar
techniques (molecular bonding) and materials (quartz and Zerodur).
Thus, the optical system consists of the following components:

\begin{enumerate}
\item
beam splitting wedge (BSW),
\item
\vspace{-8pt}
Cassegrain telescope,
\item
\vspace{-8pt}
relay lens, to achieve the required focal length, and
\item
\vspace{-8pt}
detector support.
\end{enumerate}

\subsubsection{Beam splitting wedge (BSW)}
\label{S-SIa1}

Because the wedge is the most critical element in the optical system, great
care is taken with its design and manufacture.
The wedge consists of two fused silica flats separated by an annular silica
ring polished to an angle of about 1000 arcsec.
Molecular contact bonding is used to hold the assembly in alignment.
The surfaces are flat to 1/50 wave at 630 nm and have dielectric coatings to
define the bandpass and reduce the solar transmission to an acceptable level.
The mirrored surfaces have a high reflectivity ($>0.9$) so that the intensities
of the two images are approximately in the ratio of 5:4.
The passband transmitted by the series of coatings on the wedge surfaces
is centred at approximately 615 nm and is roughly 100 nm wide.

\subsubsection{Telescope}
\label{S-SIa2}

The 17.8 cm aperture ruggedized Questar telescope is used with a reduced
aperture of 12.7 cm to accommodate the effective wedge aperture, and has a
nominal focal length of 2.5 m.
This instrument has optical elements made of fused silica and Zerodur, and a
main body and mounting parts made of Invar to minimize thermal effects.

\subsubsection{Relay lens}
\label{S-SIa3}

In order to provide the appropriate plate scale to match the resolution of
available Charge Coupled Device (CCD) detector elements, a magnification of
the image is required.
A magnification by a factor of 8, to an effective focal length of 20.5 m, is
provided by a multi-element Barlow lens.

\subsubsection{Detectors}
\label{S-SIa4}

The linear-array CCDs are mounted on a ceramic printed circuit board having a
low thermal coefficient of expansion near that of the CCD cases and a high
thermal conductivity.
This provides uniformity of expansion for the entire assembly during
temperature excursions, with the amplitude of the excursions being both small
and well characterized.
The positions of the endpoints of the CCD
elements are determined by measurement
on the Yale University microdensitometer to a precision of 2 microns.
This assembly of detectors is held in the focal plane of the instrument.
Detection of the solar images is achieved by use of seven such CCD arrays.
The central detectors are used to measure the gap, while the outer detectors
are used, in conjunction with those in the centre, to define the centres of
the images and thus the plate scale.
The linear arrays are Texas Instrument virtual phase 1728 element devices
having pixel dimensions of 12.7 x 12.7 microns and packaged in a narrow,
windowed carrier.
The CCD array is calibrated in the laboratory using an integrating sphere as
a uniform light source. A bandpass filter provides the proper wavelength of
light to match the solar input.
A geometry representative of that in the flight optical system with regard to
source diameter, angle subtended, etc. is used.
A zero-level offset and linear and quadratic sensitivity factors
are determined for each of the pixels in the seven CCDs.
This calibration procedure has produced an effective pixel uniformity of better
than 1/4\%, which is easily adequate for our present data analysis procedures.
In order to further improve the calibration, during flight we stop Sun
pointing at a given time, and let the pointing drift over the solar disk.

\subsection{Pointing System}
\label{S-SIf}

The analog pointing system, which utilizes both a LISS (Lockheed Intermediate
Sun Sensor) and real-time feedback from the detectors, achieves an operational
stability equal to or less than 12 arcsec over the entire flight (10-12 hours).

\subsection{Analog data system}
\label{S-SIg}

The function of the SDS analog data system is to control data transfer from
the CCDs to buffers that can be read by the digital data system.
Sample/Hold buffers are used to freeze the pixel signal and to set the base
level equal to the black reference level for 12 bit analog to digital
conversion.
Each of the CCDs has its own system and they are read in parallel to freeze
the image.
These conversions are performed at the (commandable) exposure rate with the
digitized values serially shifted to the 7 memory buffers in the onboard
computer at up to 4 Mbits/sec/channel. The analog data system is mounted near
the detectors to minimize noise.

\subsection{Digital data system}
\label{S-SIh}

The digital data system is based on a 66 MHz 32-bit Intel 486 processor
together with 4 Mbytes of main memory.
It uses a pair of 4 Gbyte disk drives for data and program storage, and
7 - 64Kbyte data RAM buffers.
Seven channels of serial data at up to 4Mbits/sec/channel are used to fill the
7 - 64Kbyte buffers.
A CCD clock pulse is generated at 4 times the pixel rate which is command
programmable for a rate of 4 to 20 microsec/pixel.
Detector power and control signals are supplied by the computer.
Functions of the computer are as follows:

\begin{enumerate}
\item
provide a command link between the SIP (Standard Instrument Package) and
the SDS instrument,
\item
\vspace{-8pt}
provide a data link to the ground via the SIP,
\item
\vspace{-8pt}
provide a commandable CCD pixel clock,
\item
\vspace{-8pt}
receive 7 channels of serial pixel data along with a bit and frame sync,
\item
\vspace{-8pt}
store 18 digitized data frames for each of 7 CCDs, (which we refer to as
a data cycle),
\item
\vspace{-8pt}
calculate edge data and supply computed pointing position corrections to
the onboard computer for fine pointing,
\item
\vspace{-8pt}
when pointing accuracy is satisfied, calibrate the data sets and compute
solar edge positions,
\item
\vspace{-8pt}
periodically obtain a set of data from the 7 CCDs for transmission to the
ground.
\end{enumerate}

Steps 6) and 7)
refer to in-flight determinations of the solar edges for the
purpose of fine tuning the pointing.
Section~\ref{S-DA}  describes the much more
thorough edge-determination analysis performed on the data after the flight.

\section{Observations}
\label{S-O}

Table~\ref{T-flights} lists the dates of the seven SDS balloon flights that have
yielded useful measurements of the solar diameter.
All flights were conducted by the CSBF from
their station in Fort Sumner, New Mexico.
The table lists peak altitude for each flight, which is near the float altitude
at which the observations are made.

\begin{table}
\caption{
SDS flights}
\label{T-flights}
\begin{tabular}{rcrl}
\hline
Flt. & Date      & Alt. & Comments \\
  \# &           & (km) &          \\
\hline
  6  & 1992-09-30  & 30.7 & First flight using contact-bonded wedge  \\
  7  & 1994-09-26  & 31.6 & First flight with fixed focus         \\
  8  & 1995-10-01  & 31.5 &          \\
  9  & 1996-10-10  & 32.0 &          \\
 10  & 2001-10-04  & 31.6 & Onboard memory failure; $<1$\% of data retrieved \\
 11  & 2009-10-17  & 33.3 & Anomalous excursion in image quality \\
 12  & 2011-10-15  & 31.0 &          \\
\hline
\end{tabular}
\end{table}

Flight 6 was the first to be made with the contact-bonded wedge.
A spring mechanism had been used in previous flights to try to maintain
the geometry of the wedge but this proved to be totally inadequate, thus
rendering data from the earlier flights unusable.
Between Flights 6 and 7, the SDS was sent to the White Sands Missile Range
optical instrumentation laboratory
for a thorough refurbishment;
this included making
structural changes to the primary mirror mount and locking the previously
adjustable focus in place following a careful benchtop alignment.
A failure of the onboard hard drive during Flight 10 meant that only
a small fraction of the in-flight data was recorded, as part of the
telemetry stream sent to the ground station.
Still, there is a sufficient amount of data from Flight 10 to allow for a
worthwhile radius measurement to be made.

\begin{figure}
\centerline{\includegraphics[width=0.85\textwidth,clip=]{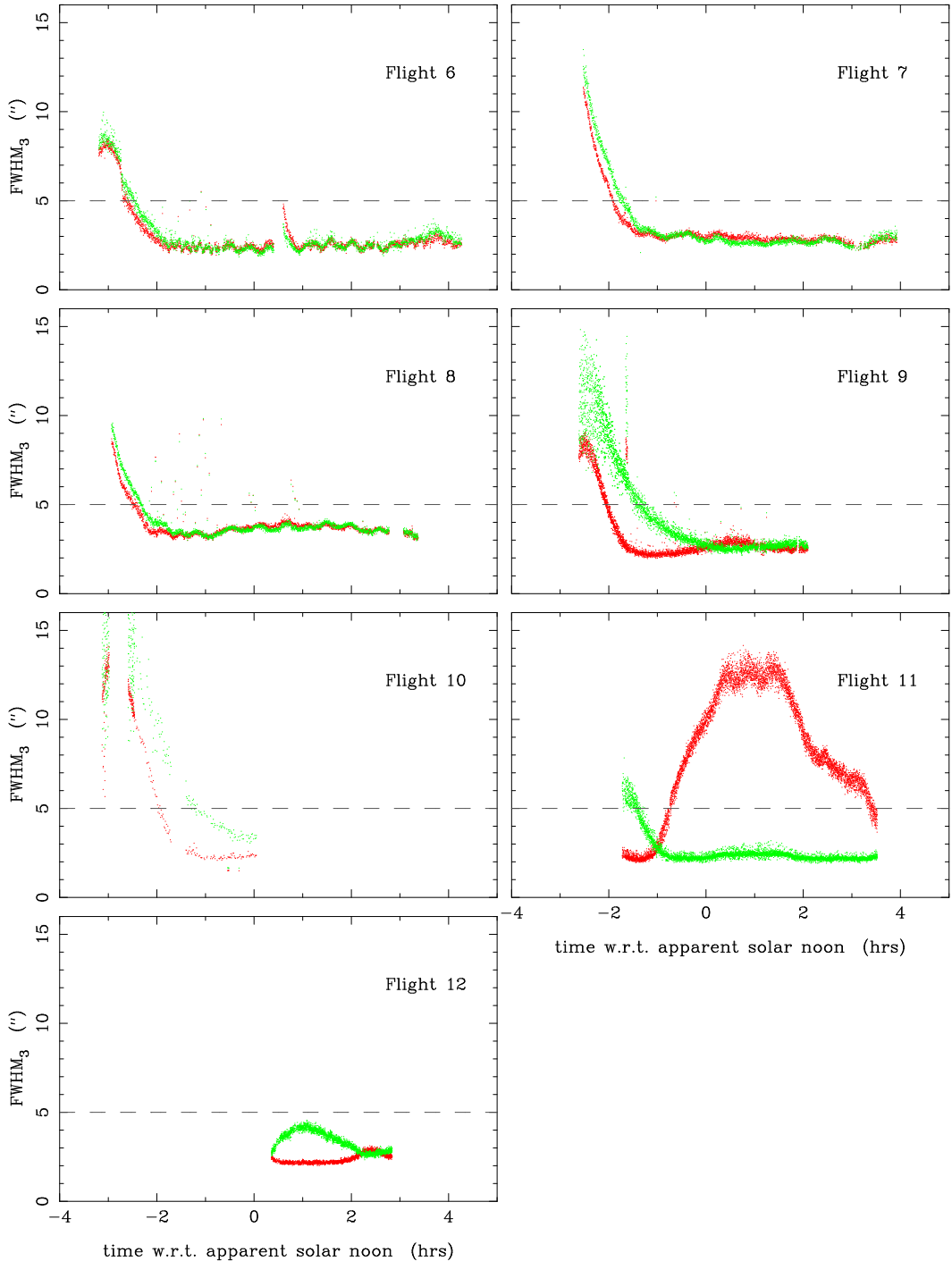}}
\caption{
Limb edge width as a function of time (relative to apparent solar noon)
throughout the seven successful SDS flights.
The FWHM of the Gaussian fit to the derivative of the smoothed profile is shown 
for the
direct image (red) and the reflected image (green) on the central CCD.
The sparse data for Flight 10 is the result of an onboard recorder failure.
The unusual behavior of the direct-image limb width during Flight 11 is
also noteworthy.
Solar diameter determinations are based
only on those portions of each flight in
which the FWHM of the direct and reflected images is relatively well-behaved.
One such criterion is that the direct-image FWHM is below 5$''$, as shown by the
dashed line.
See Section~\ref{S-DAn} for a complete description of the data trimming criteria.
}
\label{F-allwid}
\end{figure}

Flight 11 employed a substantially larger balloon than the other flights,
(29 Mcf versus 12 Mcf), accounting for its somewhat higher peak altitude.
An examination of the width of the solar limb edge (as determined using the
procedure described in Sect~\ref{S-DA}) over the course of the flight
indicates an anomalous behavior.
Figure~\ref{F-allwid}
shows the time variation of the limb width measured on the central
detector (CCD\#3), for both the direct and
reflected limb images, during all seven flights.
The initial steep portion seen in most flights is during ascent, before
thermal equilibrium is attained.
During Flight 11, in 2009, the direct-image limb widened drastically during
the float portion of the flight, while the reflected-image limb remained
steady.
The cause of this behavior is unknown; a thermally-induced bending of one
or more of the BSW surfaces might be to blame, although it would require a
strange coincidence of surface deformations to adversely affect the direct
image and yet leave the reflected image intact.
While this flight employed a larger balloon, the thermal environment for the
observations should not have differed substantially from that of
the other flights.
Regardless, because of the FWHM behavior of Flight 11, only observations near
the beginning and end of the flight will be used in our analysis.
Also, in general, it is evident that the limb-width profiles of the last
four flights are less well-behaved than the corresponding profiles of Flights
6, 7, and 8.
Possibly the instrument's optical alignment and/or rigidity were altered
after Flight 8.
It should be noted that on-board accelerometers indicate that Flight 8 was
one of several that
ended with particularly hard landings, in excess of 25 Gs.
Typically, after each flight the detector assembly is removed for photometric
calibration.
This also gives access to the instrument's optical components, which are
visually inspected, although no intentional adjustments have been made in the
optical alignment since the post-Flight 6 refurbishment at White Sands.
While the properties of the instrument are subject to change over time, the
relative angles of the surfaces in the optically bonded beam splitting wedge
are expected to remain constant.

\section{Data Analysis}
\label{S-DA}

An SDS flight typically consists of hundreds of thousands of individual
exposures, each exposure imaging the Sun's limb at ten different position
angles, in total, across the seven CCDs (seen in Figure~\ref{F-layout}).
The $7$-$\mu$sec duration exposures are grouped into ``cycles'' consisting of 18
exposures per cycle.
The 18 exposures are completed within 0.22 sec
and the time between successive
cycles ranges from roughly 1 to 3 sec, depending on
the flight.
Pixel data from each cycle (7 CCDs times 18 exposures) are
stored together in a single file, along with housekeeping data for the
instrument at the time of observation.
The post-flight processing pipeline treats a cycle's worth of data at a time,
although each exposure within a cycle is, for the most part,
processed independently.
Only in the final step (a correction based on the difference in width
of opposite limbs of the Sun and its variation throughout the flight)
does information from one cycle affect the analysis of another.

The sequence of steps in the processing pipeline is listed below and a
detailed description of each step is given in the subsections that follow:

\vspace{10pt}
\indent
 1. adjust photometrically, i.e., flatfield \\
\indent
 2. remove ``spike'' artefacts \\
\indent
 3. perform preliminary edge detection for all ten limbs \\
\indent
 4. subtract background level and ``ghost'' image \\
\indent
 5. make final edge determinations for the limbs using the cleaned profiles \\
\indent
 6. correct for bias due to instrumental broadening of the limbs \\
\indent
 7. transform CCD edge coordinates into focal-plane $(x,y)$ \\
\indent
 8. correct $(x,y)$ positions for optical distortion \\
\indent
 9. correct $(x,y)$ positions for atmospheric refraction \\
\indent
10. fit direct and reflected image measures to two circular arcs \\
\indent
11. determine the minimum ``gap'' between direct and reflected limbs \\
\indent
12. calculate $R_{sun}$ in terms of the BSW angle \\
\indent
13. correct for Earth/Sun distance \\
\indent
14. detrend $R_{sun}$ as a function of delta limb width for the entire flight \\

\subsection{Photometric adjustment}
\label{S-DAa}

Prior to most flights, the response of the seven TC101 linear-array
CCD detectors is calibrated from a series of benchtop observations of a
reference sphere.
A sequence of exposures of varying duration sample the full dynamic
range of the detectors.
From these measures, a quadratic function is fit to the CCD reading as a
function of exposure time, on a per pixel basis.
This is essentially a photometric bias plus nonlinear flatfield correction
for each detector.
Such calibration coefficients were determined prior to flights 6, 7, 8,
9 and 11.
In some cases, all coefficients were determined in full (Flights 6 and 8)
while in other cases only the constant (Flights 7 and 8)
or constant and linear terms were redetermined (Flight 11)
from the benchtop measurements.

In the data processing pipeline,
each flight's data are reduced using the corresponding pre-flight coefficient
sets, with the exception of flights 10 and 11; the processing of these two
flights makes use of the pre-flight 8 coefficients.

\begin{figure}
\centerline{\includegraphics[width=0.8\textwidth,clip=]{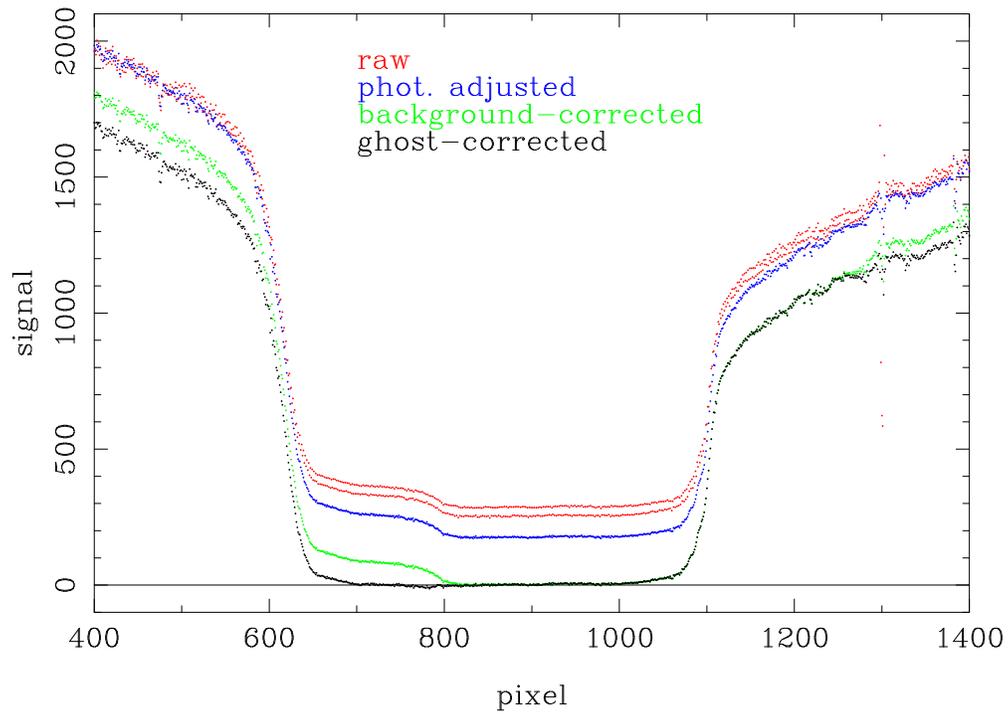}}
\caption{
A sample SDS limb profile, obtained with the central CCD, at various stages
of the reduction pipeline.
The figure shows the raw profile (red); after photometric adjustment (blue);
after background-level correction (green); and after the ``ghost'' image
is subtracted (black).
See the text for a description of the different correction steps.
}
\label{F-unghost}
\end{figure}

Subsequent to the application of these second-order coefficients,
an additional photometric correction is needed.
The TC101 detectors incorporate two amplifiers per detector,
controlling separately the odd- and even-numbered pixels.
The benchtop calibration sequences cannot account for differences in the
bias levels between the two amplifiers as these change every time the
electronics are powered.
Thus, a scalar adjustment is calculated based on the empirical differences
between neighboring even and odd pixel values, effectively putting all pixels
on the photometric system of the odd pixels.
The seven scalar corrections, one per detector, are calculated separately
for each exposure within the flight.
Figure~\ref{F-unghost}
presents a sample profile of the central detector, before (red points)
and after (blue points) the photometric adjustments are made.
What appears to be two curves in the raw data is simply the result of the
odd/even pixel offset.

\subsection{Spike removal}
\label{S-DAb}

Following the flatfield and bias corrections, the observed image profiles
contain noise in the form of spikes that are one to several pixels wide
and of relatively large amplitude.
These spikes can fool the subsequent edge-detection algorithm, which relies on
searching the numerical first derivative of the profile.
For this reason, the profiles are filtered by examining the difference in
value between each pixel and the average value of the pixels that fall
4 pixels to either side.
Any pixel $i$ whose value differs by more than 120 counts has its value replaced
by the $(i+4,i-4)$ pixels' average.
Note that the CCD gain is$\sim$8 e-/count and the 120-count spike threshold
is 3 to 4 times the standard deviation of the signal for the portion of the
CCDs containing the sun's image.
In practice, 0.1 \% of the pixels are tagged as belonging to a spike.
The spike-removal procedure effectively eliminates the problem of false
edge detections.

\subsection{Preliminary edge detection}
\label{S-DAc}

We adopt as the conceptual limb edge the inflection point of the profile.
This is determined by finding the maximum (or minimum, depending on whether
the edge is a rising or falling one) in the first derivative of the profile.
Unfortunately, the numerical derivative is noisy, creating local maxima and
minima that confuse the edge detection routine.
Thus, in practice, the profile is smoothed before differentiation.
We use 20 repeated applications of a 3-pt smoothing.
This necessary smoothing will unavoidably broaden the limb and,
in doing so, will shift the inflection point.
The shift, or bias, is a consequence of the difference in profile shape just
before and after the inflection point.
This bias will be corrected in a subsequent reduction step.
Note that even absent the artificial smoothing introduced at this
step in the reduction, other factors (primarily the instrument response)
have already broadened the limb significantly.
No matter the cause of the broadening, it will result in a bias in the
measured inflection point position.
The correction is calibrated as a function of the overall broadening
and thus will correct for the overall bias, i.e., both the unavoidable component
due to the instrument and that due to the 3-pt smoothing.

\begin{figure}
\centerline{\includegraphics[width=0.8\textwidth,clip=]{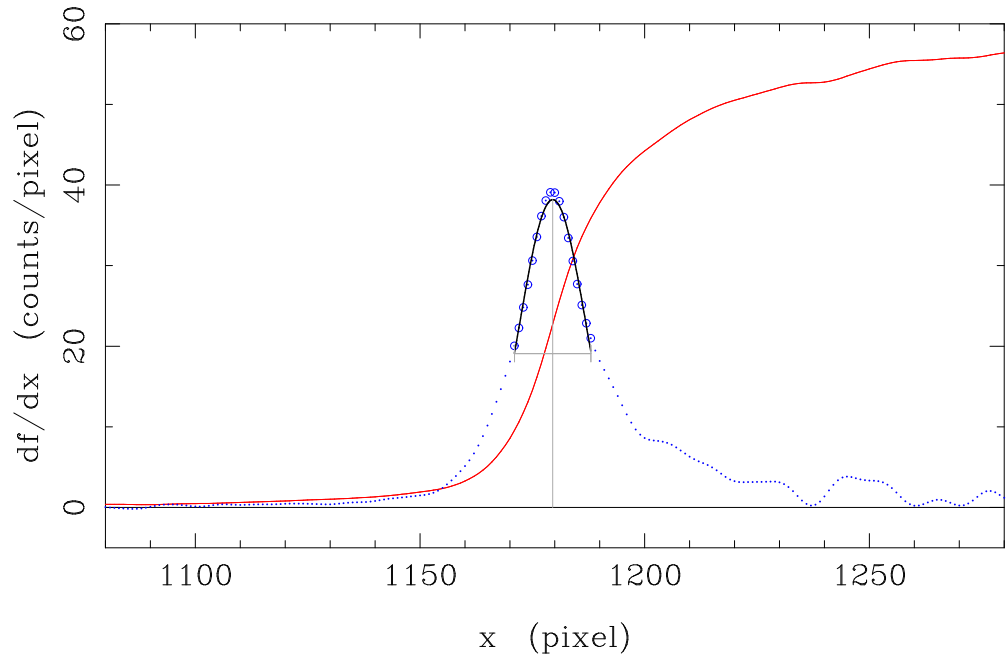}}
\caption{
Illustration of the technique used to determine the inflection point
position, $IPP$.
The red curve is the smoothed limb profile.
Shown in blue is the numerical first derivative of the profile.
The heavy blue symbols indicate those points used in the least-squares
Gaussian fit, which is shown by the black curve.
The centre and FWHM of the fitted Gaussian are indicated by the
gray line segments.
}
\label{F-wingsplot}
\end{figure}

Once smoothed, a simple numerical derivative is obtained,
$f'_i = (f_{i+1}-f_{i-1})/2$.
The index of the maximum value of $f'_i$ is identified, and all adjacent
points for which $f'$ is greater than 0.5 times the maximum value are used
to better refine the inflection point location.
This refinement is by way of a Gaussian fit to the extracted points.
The centre of the Gaussian is taken as the inflection point position.
The FWHM of the best-fitting Gaussian is also stored.
See
Figure~\ref{F-wingsplot}
for an illustration of the edge-fitting procedure.
In this manner, preliminary positions for the ten limb edges are found,
in terms of pixel position along the seven linear-array detectors.

\subsection{Background and ghost image subtraction}
\label{S-DAd}
A careful examination of Figure~\ref{F-unghost}
reveals the presence of a low-amplitude
``ghost'' of the main falling limb, lying in the valley between the falling
(direct) and rising (reflected) limb images and offset from the main
image by about 174 pixels.
This image is due to a pair of internal reflections within a single element
of the BSW.
(The front and back surfaces of the two ``flats'' from which the BSW is
composed are slightly non-parallel, to avoid fringing.)
It is important to subtract this ghost image from the observed profile
since its presence can affect the inflection point position, ($IPP$);
primarily that
of the falling limb because of the direction of the ghost-image offset.
A ray-tracing model of the as-designed SDS optics predicts an offset
of 178 pixels for the ghost image.
Empirically, it is found to lie at an offset of 174 pixels.
Its amplitude is a less predictable function of the coatings on the relevant
surfaces of the BSW; coatings that deteriorate over time.
For this reason, the relative amplitude of the ghost image is determined
for each exposure, while the offset - which depends only on the assumed
invariant geometry of the wedge - is set at 174 pixels.

Note that the rising-limb profile also has a contamination from its ghost,
although it is difficult to detect as the direction of the offset causes
it to fall entirely within the high-signal portion of that limb.
Still, for the sake of consistency, this limb should also have the ghost
profile removed.

In order to subtract the ghost-image profile properly,
the relative amplitude of the direct and reflected limbs must be
determined, as well as the relative amplitude of the ghost and main images.
This is accomplished by measuring the amplitude of the central detector's
observed profile at key points with respect to the preliminary inflection
point positions of the main limb edges, $IPP_D$ and $IPP_R$,
as derived in the previous step.

The observed profiles in all of the SDS detectors are elevated by a
background ``sky'' level that is a combination of true scattered light
and uncorrected electronic bias, (presumably dominated by the latter).
The background level is taken to be the minimum value of an 11-pixel-wide
moving average within a conservatively limited range of pixels presumed
to be free of either (direct or reflected) solar limb and their ghost images.
The pixel ranges are based on the preliminary edge positions, the preliminary
FWHM limb edge width, the 174-pixel ghost offset, and an additional cushion
of 30 pixels.
Thus, for detectors 6 and 7, which contain only the direct image limb on
the low-pixel end of the detector,
the signal-free pixel range is from the preliminary inflection point position
$(IPP_D)$ plus the FWHM plus the cushion up to 1728 minus the cushion.
For detectors 1 and 5, which contain only the reflected image limb on the
high-pixel end of the detector,
the relevant range is from the cushion value up to the preliminary
$IPP_R$ minus the FWHM minus the cushion.
Finally, for detectors 2, 3 and 4, which contain both direct and reflected
image limbs, the signal-free pixel range is from the preliminary $IPP_D$
plus the 174-pixel ghost offset plus the cushion and extending up to
the preliminary $IPP_R$ minus the FWHM minus the cushion.
Figure~\ref{F-unghost}
shows a sample profile from detector \#3 after subtraction of the
background level calculated in this manner.

\begin{figure}
\centerline{\includegraphics[height=0.7\textwidth,clip=,angle=-90]{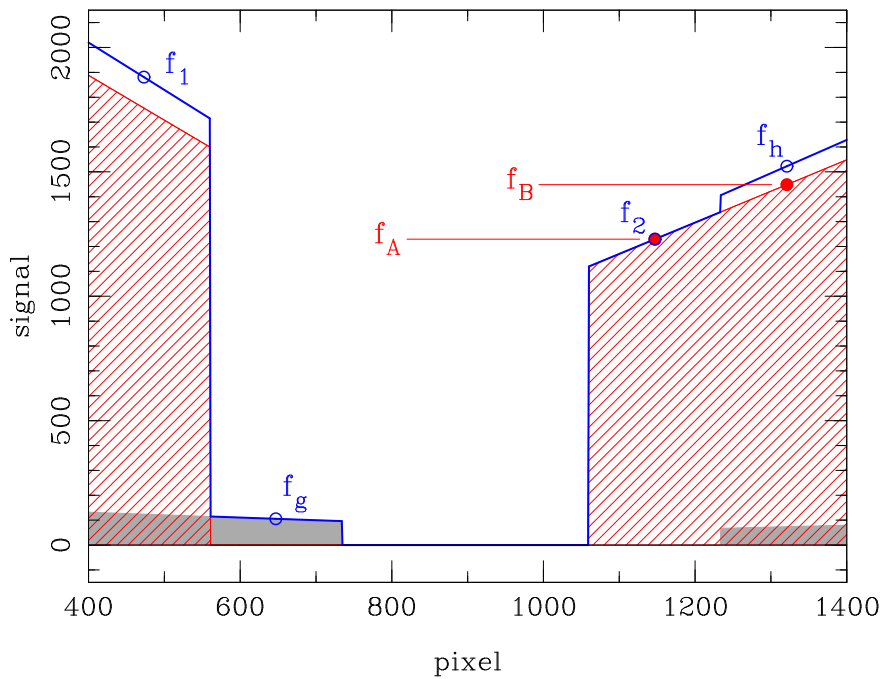}}
\caption{
An idealized representation of the direct and reflected solar-limb profiles,
as seen by the central detector, and their decomposition into ``main-'' and
``ghost-''image contributions.
The main images are shown in red-hatch shading; the direct-image limb to the
left and the reflected-image limb to the right.
Shaded in gray is the ghost image of the main image, which is caused by
internal reflection within a single element of the beam splitting wedge.
The ghost image is offset from the main image by 174 pixels and is
greatly reduced in intensity.
The main and ghost images combine to produce the observed total profile,
represented by the
heavy blue curve.
By measuring the amplitude of the observed profile at key points relative to
the direct- and reflected-limb edges (located at $\sim$560 and $\sim$1060,
respectively, in the example shown), it is possible to deduce the relative
intensity of the direct image to that of the reflected image, and the
relative intensity of the ghost image to that of the main image.
}
\label{F-fourpts}
\end{figure}

Having subtracted the background level, it is now possible to calculate the
relative amplitudes of the various limb images - reflected relative to direct
and ghost relative to main.
This is accomplished by measuring the height of the profile in the central
detector \#3 at four key positions, i.e., just above the ``knee'' relative to
the main direct and main reflected limb edges, and at similar
pixel offsets relative
to the projected positions of their respective ghosts.
See Figure~\ref{F-fourpts}
for a schematic representation of these four key points.
A simplified geometry for the underlying instrumental limb is shown for the
sake of clarity, but the strategy is valid for the actual observed limb.

The offset from the $IPP$ of a limb edge and what is meant by
``just above the knee'' is largely arbitrary; we take its value to be one half
the ghost image offset, or $\Delta_{knee}$=87 pixels.
Thus, the needed profile amplitudes are \\

\indent
 $f_1 \equiv f(x_1)$ where $x_1=IPP_D-\Delta_{knee}$ \\
\indent
 $f_g \equiv f(x_g)$ where $x_g=IPP_D+\Delta_{ghost}-\Delta_{knee}$ \\
\indent
 $f_2 \equiv f(x_2)$ where $x_2=IPP_R+\Delta_{knee}$ \\
\indent
 $f_h \equiv f(x_h)$ where $x_h=IPP_R+\Delta_{ghost}+\Delta_{knee}$ \\

Determination of the various $f_i$ is made by a linear fit to the observed
profile at the pixel nearest $x_i$ along with its three neighbors on
either side, and then evaluation of the fit at $x_i$.
With $f_1, f_g, f_2,$ and $f_h$ in hand,
the relative amplitudes are derived in the following manner.

Assume the relative amplitude of the reflected and direct images is $\Gamma$,
and the amplitude of the ghost profile relative to the main profile is
$\gamma$.
The observed, central-detector profile can be thought of as being built up
from an intrinsic instrumental solar limb profile that has added to it a
reflected, offset and scaled copy of itself, and then this combination having
added to it a ghost copy of itself, with a separate offset and scaling factor.
(In reality, there are second- and higher-order ghost images also present,
but each of these is successively diminished by the factor $\gamma$ and in
practice only the first-order ghost need be considered.)

For convenience, our formulation actually uses the main reflected image as
the base template from which the overall profile is to be
constructed.
Adopting this, there are two points in the template profile that are
useful to the formulation; call these $f_A$ and $f_B$, as shown in
Figure~\ref{F-fourpts}.
The four points measured in the observed profile can be written in terms
of these two points' amplitudes and the relative scale factors,

\begin{equation}
\begin{array}{l}
 f_1 = f_A/\Gamma + \gamma f_B/\Gamma \\
 f_g = \gamma f_A/\Gamma \\
 f_2 = f_A \\
 f_h = f_B + \gamma f_A
\end{array}
\end{equation}

These can be combined to isolate the reflected-to-direct relative scale,
$\Gamma$,

\begin{equation}
 0 = (f_g f_g/f_2)\Gamma^2 + (f_1 - f_g f_h/f_2) \Gamma - f_2.
\end{equation}

Having found $\Gamma$,
the ghost-to-main image relative scale, $\gamma$, can be determined,

\begin{equation}
 \gamma = (f_g/f_2)\Gamma.
\end{equation}

\begin{figure}
\centerline{\includegraphics[width=0.8\textwidth,clip=]{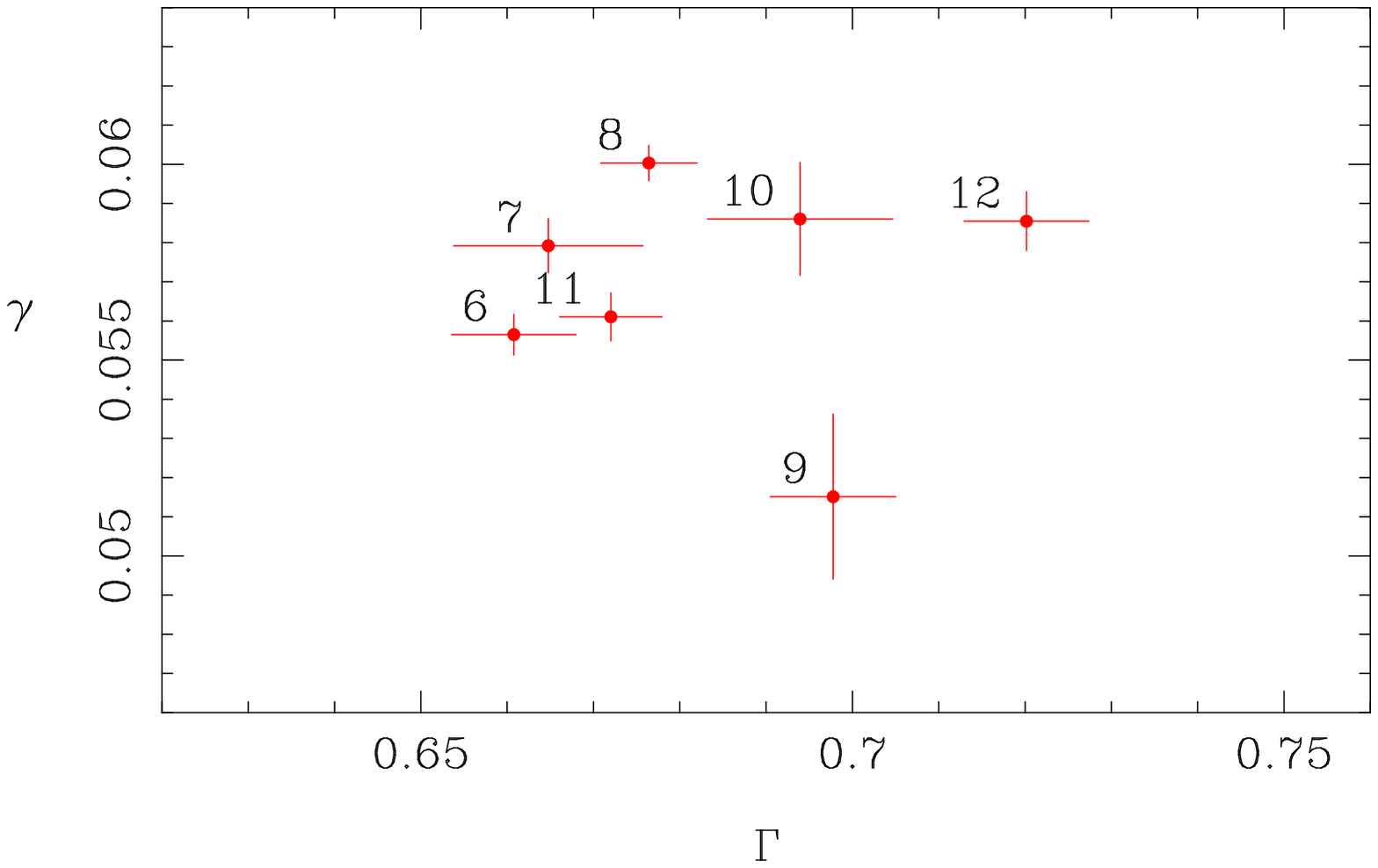}}
\caption{
Ratio of the ghost-image amplitude to that of the main image ($\gamma$)
versus the ratio of the reflected-image amplitude to that of the direct
image ($\Gamma$).
The data points are labelled by flight number.
The ``error'' bars show the rms of the estimated ratios within each flight.
}
\label{F-gamma}
\end{figure}

Typical values of $\Gamma$ and $\gamma$ for the SDS flight data are
approximately 0.7 and 0.06, respectively.
These vary somewhat from flight to flight, as shown in Figure~\ref{F-gamma};
presumably, this
is due to aging of the surface coatings of the beam splitting wedge.
This is discussed more thoroughly in Section~\ref{S-D}.

From the ghost scale factor, $\gamma$, and the assumed
$\Delta_{ghost}=174$ pixel, it is
possible to build a ghost-free template from the reflected-image portion
of CCD\#3's profile.
Our pipeline constructs one template per cycle from the average of the 18
exposures, interpolated and stored at a super-resolution of 0.1 pixels.
Knowing the relative scale factors, ($\Gamma$ and $\gamma$),
the preliminary main edge
positions, ($IPP_D$ and $IPP_R$), and adopting a ghost-image offset of
$\Delta_{ghost}=174$
pixels along CCD\#3, the template can be used to subtract the ghost image
from all the detectors' profiles.
Note that the ghost-image offset for the outer detectors, in pixels, is
larger by a factor of $1/\cos(\theta)$, where $\theta=27.1^{\circ}$
is the angle between
the outer detectors and the axis of deflection of the beam splitting wedge,
which is approximately aligned with the inner detectors.
Additionally, a pixel scale factor must also be applied to the template
when used on the outer detectors to account for the slight misalignment of
a radial for the main image and that of the ghost image.
The misalignment is roughly $\phi=1.3^{\circ}$, and the correction factor
is $\cos(\phi)$.

Using the above procedure, the ghost image can be subtracted from the
observed profile in each of the seven detectors.
The result for our sample CCD\#3 profile is shown in Figure~\ref{F-unghost}.

\subsection{Final edge determination}
\label{S-DAe}

With all pre-processing completed, including the subtraction of the ghost
images, the inflection point positions ($IPP$s) are once again determined.
This is done as before, using the Gaussian fit to the numerical first derivative
procedure outlined in Section~\ref{S-DAc}.
The result is ten $IPP$s per exposure, along with accompanying FWHM measures.

\subsection{Limb-broadening bias}
\label{S-DAf}

\begin{figure}
\centerline{\includegraphics[width=0.8\textwidth,clip=]{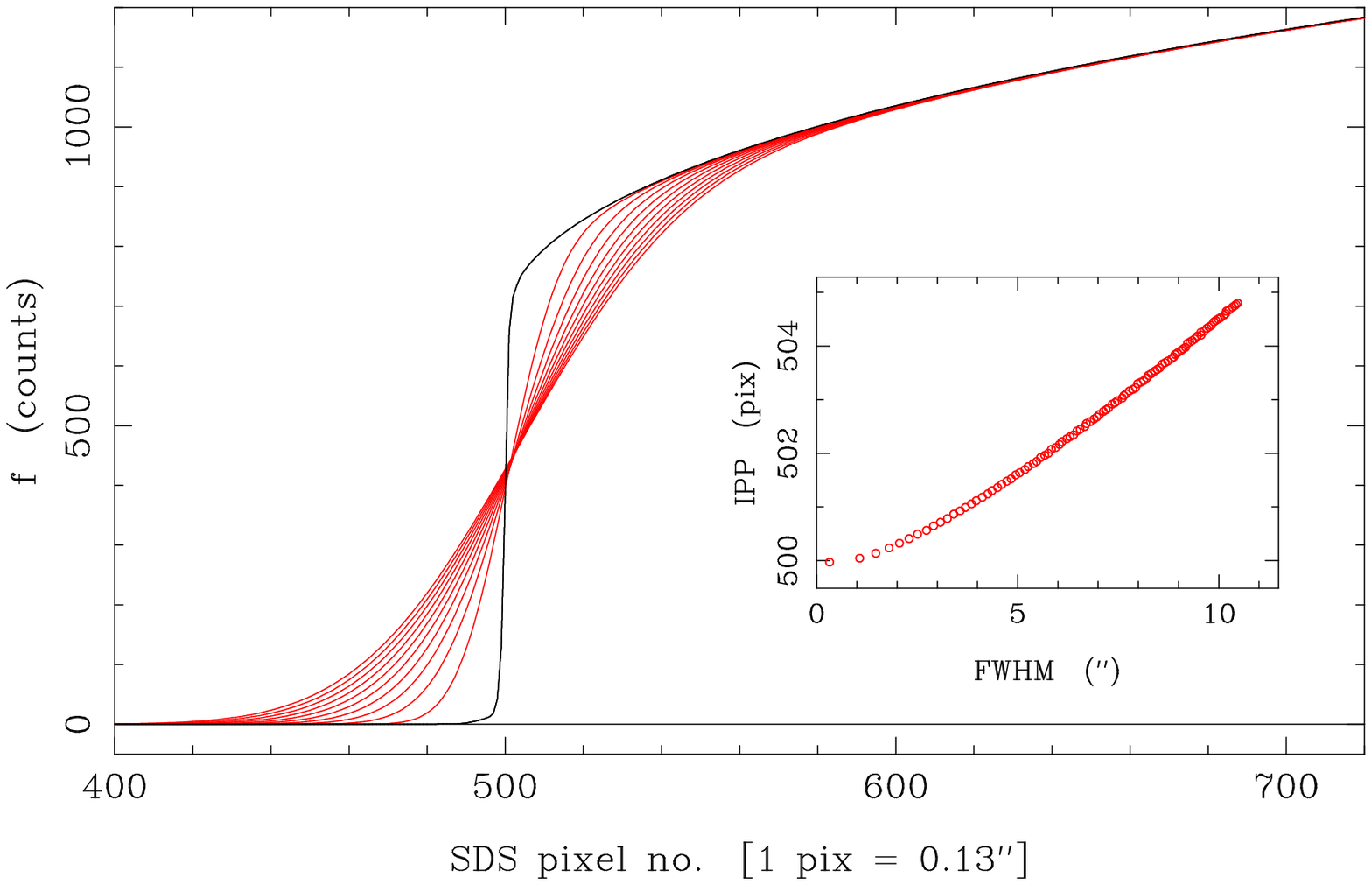}}
\caption{
Calibration of the limb-broadening bias correction.
The black curve displays the synthetic limb constructed from the COSI
model mated to a Hestroffer \& Magnan type profile, arbitrarily positioned
at an SDS pixel value of 500.
The red curves indicate the effect of successive smoothing (broadening)
of the original profile.
Using the method described in the text to determine the inflection point
position ($IPP$), i.e., Gaussian fit to the profile's first derivative, the
shift in $IPP$ is a monotonic function of the limb width, as shown in the
inset panel.
}
\label{F-fitsim}
\end{figure}

The shape of the limb edge is asymmetrical about the inflection point.
For this reason, any mechanism that broadens the limb will impart unequal
amounts of influence from either side of the $IPP$ and, thus, will shift it.
One expects (and finds) a one-to-one correspondence between limb width and the
net amount of shift or bias.
A calibration of this broadening bias can be made using a sufficiently
accurate synthetic limb profile.
The Code for Solar Irradiance
(COSI, Haberreiter et al.~2008; Shapiro et al.~2010)
calculates the solar spectral irradiance and has been used to calculate
the intrinsic solar limb profile for the passband of the SDS.
The profile provided by COSI was of sufficient resolution near the limb edge,
but lacking toward the interior of the solar disk.
For this reason, we have mated the COSI profile to the analytical solar
limb prescription given by
Hestroffer \& Magnan (1998).
The resulting synthetic profile is shown in Figure~\ref{F-fitsim},
scaled to nominal SDS pixels.

Applying the method of Section~\ref{S-DAc}
(that of Gaussian fitting of the first
derivative of the profile) provides the $IPP$ and FWHM for the intrinsic limb.
The synthetic profile can then be broadened by a specified amount and the
fitting procedure repeated, noting the shift in $IPP$ from that of the
unbroadened profile.
The $IPP$ shifts are plotted versus measured limb broadening in the inset
panel of Figure~\ref{F-fitsim}.
Two different forms of broadening were explored -- multiple applications of simple
three-point smoothing, and convolution with a Gaussian kernel.
The resulting $IPP$-shift curve was practically identical for the two forms
of broadening, being characterized only by
the FWHM measure of the broadened limb.
A best-fitting polynomial description of this calibration curve is incorporated
into the SDS processing pipeline.
The correction is calculated as the difference between the value of the
calibration curve at FWHM=0.32 arcsec (the width of the unbroadened COSI
profile) and its value at the measured FWHM of the limb edge being corrected.
A FWHM-based bias correction is applied to the $IPP$ values of every
limb-edge determination.

\subsection{Transformation to Cartesian coordinates}
\label{S-DAg}

The $IPP$s of the ten limb edges are each a one-dimensional measure of position
along a linear-array CCD.
These positions are transformed into focal-plane cartesian coordinates to
facilitate circle-fitting of the direct and reflected disk images.
The CCDs are soldered onto a ceramic mounting block, defining their orientations
in the focal plane.
The endpoints of the seven CCDs, i.e., the midlength points of the
outer edges of the first and last active pixels, were measured in 1988 using
the Yale PDS microdensitometer in opaque light-source mode.
The repeatability of the endpoint measures was found to be $\sim2$ microns.
(The Yale PDS is a submicron-accuracy machine; difficulty in defining the 
low-contrast edges of
the active areas of the SDS detectors was responsible for the 2-micron
precision.)

Note that while 2 microns in the SDS focal plane corresponds to 20 mas, the
geometric arrangement of the detectors - primarily the fact that the
``gap'' between direct and reflected image edges is measured within
individual detectors - lessens the sensitivity of the derived solar radius
to uncertainties in the detector positions.
Monte-Carlo type tests in which the detectors were randomly offset and       
rotated by amounts consistent with the 2-micron uncertainty in their endpoint 
positions yielded just 3.5-mas variation (rms) in the resulting radius
determinations.
This amount, in quadrature, has little influence on what will be shown to be
an overall 20-mas systematic error estimate for the SDS radius measures.

Concerned that numerous hard landings after balloon flights might have
jarred the detector enough to shift the relative positions of the CCDs,
the detector assembly was again measured in early 2012 using an OGP Avant 600
measuring machine also at Yale.
The agreement between the 1988 PDS measures and the 2012 OGP ones is at an
rms level of 4 microns, the expected accuracy of the OGP.
Thus, it is reasonable to assume that the physical locations of the CCDs have 
remained stable over the course of the SDS balloon flights.
The 1988 PDS endpoint measures are adopted for our analysis.

A simple linear interpolation of the $IPP$, in pixels, and the known $(x,y)$
positions of the CCDs' endpoints provides the necessary transformation.
Note that the detector assembly is typically unmounted and remounted between
flights and an effort is made to align the central CCD with the wedge angle
plane, such that the direction of offset of the reflected image coincides
with CCD\#3, the central detector.
In practice, the alignment is not perfect and the circle-fitting procedure
described in Section~\ref{S-DAj}
allows us to determine the slight misalignment.
The actual alignment is taken into consideration when determining the
separation and minimum gap of the two fitted circles.

\subsection{Optical distortion}
\label{S-DAh}

Optical distortion of the as-designed SDS instrument has been determined
independently using two different ray tracing programs.
These agree in their finding that the maximum distortion occurs approximately
40 mm from the centre of the focal plane, with an amplitude of 0.018 mm.
As discussed in the description of Step 10, the apparent radius of the direct
disk image can differ from that of the reflected disk image, in general.
One possible explanation for this asymmetry is an optical field-angle
distortion (OFAD) that is not radially symmetric with respect to the centre
of the detector plane.
We have explored this possibility in two separate ways; by adopting the
design OFAD but having it offset from the centre of the detector plane, and
by using a ray tracing program to calculate the OFAD under the assumption
that the Barlow lens is tilted by up to several degrees.

The first approach allows a reconciliation of the direct and reflected radii
for Flights 6 through 8, given OFAD centre offsets of tens of mm.
The ratio of radii is more deviant in Flights 9 through 12,
($R_{sunD}/R_{sunR}$=0.992);
no amount of centre shift can bring the ratio to unity.
However, the second approach, that of introducing a tilt in the Barlow lens,
does allow the ratio to be forced to unity for all flights, using three
different values of tilt, although the tilt angle exceeds $10^{\circ}$ for the
later flights, an unrealistically large value.
Still, it is informative that even under such extreme conditions of tilt and
OFAD centre shifts, the effect on the resulting $R_{sun}$ values is slight.

A series of reductions of the SDS flight data have been made, using a large
range of OFAD models and corrections.
These sets of reductions are used to estimate the possible systematic errors
in the final $R_{sun}$ values.

\subsection{Atmospheric refraction}
\label{S-DAi}

Even at the SDS float altitude (atmospheric pressure of 3 - 5 mbar), the
instrument observes through some residual atmosphere and correction must be
made for the effect of differential refraction over an angle corresponding
to the size of the Sun.
The formulation adopted is based on that of
Smart (1979)
and is described in detail by
Sofia et al.~(1994).
In practice, the correction ranges from about 5 to 25 mas in the vertical
direction and remains less than 5 mas in the horizontal direction.

\subsection{Circle fitting}
\label{S-DAj}

Having been corrected for distortion and refraction, the limb shape is taken
to be circular; the solar oblateness being ignored at this point.
The $IPP$s from a single exposure
provide five points along the limb of the direct image and another five
points along the limb of the reflected image.
For each of the two disk images the points span an arc of about $54^{\circ}$,
sufficient to define the circles.
Each circle, uniquely specified by its radius and the $(x,y)$ location of
its centre, is overdetermined by the five measures.
The ``geometric'' best fitting circle to a set of points minimizes the
sum of the squares of the distances of the points from the circle.
This is a nonlinear least-squares problem, in general, but a numerically
adequate and computationally fast approximate solution is available to us,
given the particular layout of the SDS measures.
The location of the two outer points, at roughly $\pm27^{\circ}$ relative to
the central concentration of three CCDs, naturally give the outer points a
more prominent role in constraining the best-fitting circle.
We adopt as an approximation to the geometric best-fit circle, the mean of
the three distinct circles that are defined by the two outer points and
each of the three inner points, in turn.
This procedure assigns a slightly higher weight than deserved to the outer
points, but, in practice, the random error introduced by this expedient
approximation is negligible.
In this manner, each SDS exposure yields radius and centre
measures for both the direct and reflected disk images.
The separation of the two centres determines the instantaneous pixel
scale for the exposure.
The two radii provide a helpful diagnostic and constraint on the form
of the distortion correction, but these do not directly enter into the
determination of $R_{sun}$.
Instead, this derives from Equation 4 (in Section~\ref{S-DAl})
and the measurement of the minimum
gap between the two disk images.

\subsection{Minimum gap determination}
\label{S-DAk}

The minimum gap between the direct and reflected images is found from
two separate quadratic functions describing $x$ as a function of $y$ for
the three direct-image limb positions and the three reflected-image limb
positions measured by the central CCDs.
The gap is first measured in the coordinate system defined by CCD\#3 and then
corrected for the slight angle between this detector and the
direction of displacement of the
BSW, (as measured by the relative displacement of the direct
and reflected disk image centres).

\subsection{Calculation of $R_{sun}$}
\label{S-DAl}

Referring to Figure~\ref{F-layout},
the angular half-diameter of the solar disk is

\begin{equation}
R_{sun} = W (1-d/D)
\end{equation}
where $W$ is the wedge angle, nominally taken to be 989.47 arcsec, and
$d$ and $D$ represent the gap and separation.

\subsection{Correction for Earth/Sun distance}
\label{S-DAm}

The SDS flights are made during northern-hemisphere fall, at which time
the Earth/Sun distance is changing rapidly.
It is a simple matter to correct the $R_{sun}$ measures within any flight to
their corresponding values at 1 AU based on the well-known ephemeris of
the Earth's orbit.
The distance at time of observation, in AU, functions as a simple scaling
factor for the measured $R_{sun}$ values.

\subsection{Dependence on direct/reflected limb-width difference}
\label{S-DAn}

\begin{figure}
\centerline{\includegraphics[width=0.95\textwidth,clip=]{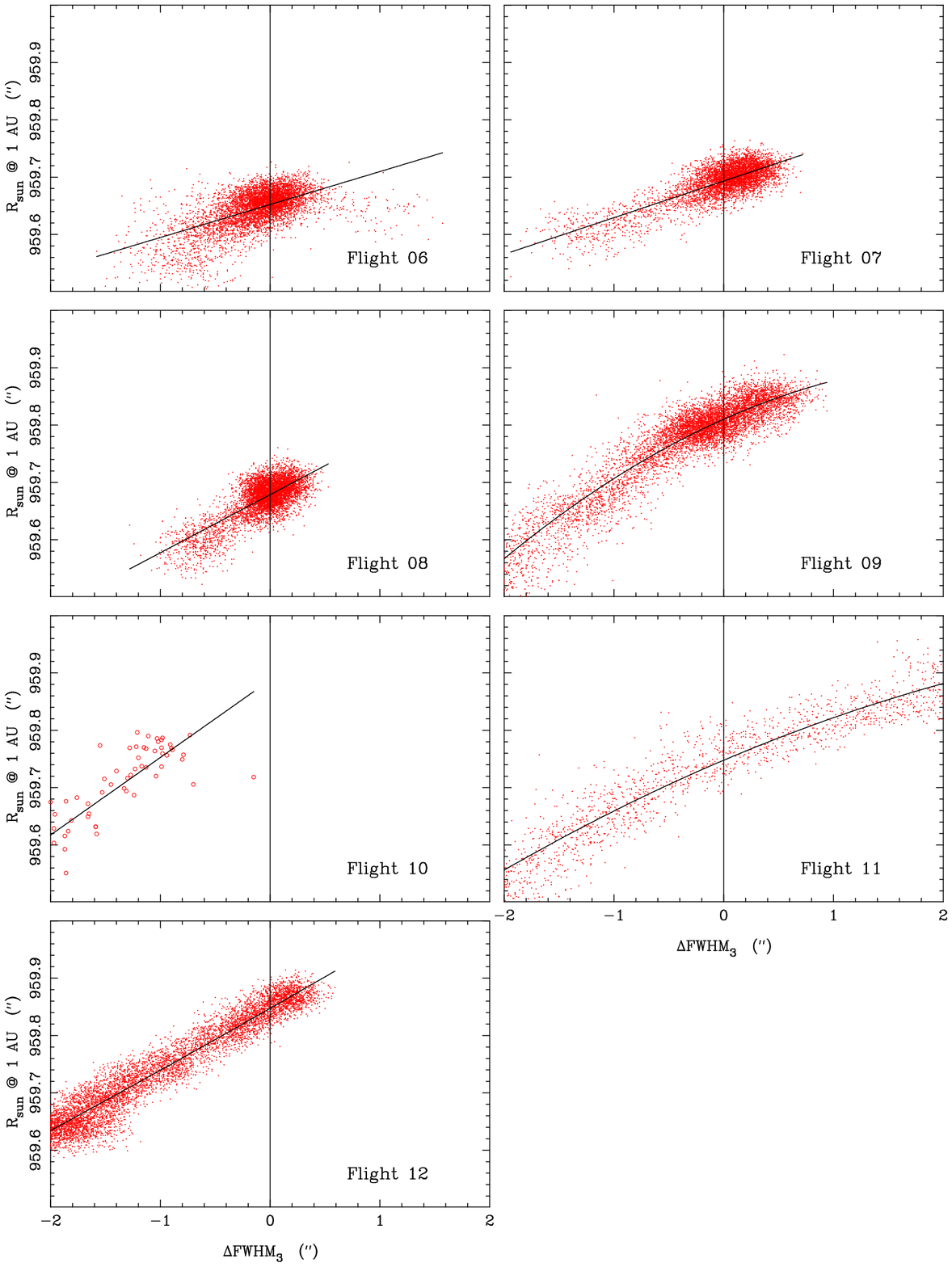}}
\caption{
Measured solar half-diameter (corrected to 1 AU) as a function of difference
in width of the direct- and reflected-image limb profiles on the central
CCD detector.
The linear or quadratic fits shown are applied to the flight data as a
correction for the observed trend.
Only data within the fitted range ($-2'' < \Delta$FWHM $ < 2''$) are valid for
correction; data outside this range are discarded.
This results in the rejection of a large portion of Flight 11 data, as can
be inferred from Figure~\ref{F-allwid}.
}
\label{F-delwid}
\end{figure}

Despite the use of low-thermal expansion materials in the telescope assembly,
the reliance on the stability of the BSW reference angle, and the
correction of the $IPP$s for the in-flight variation of the measured limb
widths, the resultant $R_{sun}$ measures tend to vary significantly throughout
the cruise portions of the SDS flights.
The cause of this variation is the differential behavior of the
direct and reflected images throughout the flights, which is discussed further
in Section~\ref{S-D}.
The SDS operating concept is based on the direct and reflected disk images
differing only by their being offset by the BSW angle.
And yet, the direct and reflected images' observed properties do differ as
witnessed by the measured widths of their respective limb images, (see
Figure~\ref{F-allwid}).
Thus, a final, crucial correction must be made for the difference in the
two image systems, one parametrized by the difference in limb widths.
Figure~\ref{F-delwid}
shows the strong correlation between measured $R_{sun}$ value and
$\Delta $FWHM$_3$, the difference in measured FWHM of the direct and reflected
limbs in the central CCD\#3.
The measures to be trusted are those in which the instrument performs as
designed, with the direct and reflected images being similar in their
properties, i.e., at $\Delta $FWHM$_3=0$.
In practice, instead of limiting ourselves to the relatively small number of 
exposures for which this holds absolutely true, we limit the exposures to 
those with $-2'' < \Delta$FWHM$_3 < 2''$ and adjust for the well-determined 
trends.
Additionally, only data from each flight in which the central-CCD direct image
FWHM $<$ 5 arcsec are used to calculate the solar diameter.
This effectively discards from the analysis
observations during the initial portion of each flight, prior to thermal
equilibrium being reached, and the anomalous portion of Flight 11.

A suspected mechanism for the observed correlation with limb-width difference
is asymmetry of the derivative profile, beyond that of the intrinsic solar
limb profile.
This has been explored, as detailed in Appendix A.
While a portion of the in-flight variation can be 
explained as being due to instrument-induced skew in the derivative profiles,
there is a significant portion of the variation that does not correlate with
skew.
Empirically, the measured diameter variation correlates best with $\Delta$FWHM
and it is this correlation that we use to make the necessary correction.

\begin{figure}
\centerline{\includegraphics[height=0.8\textwidth,clip=,angle=-90]
{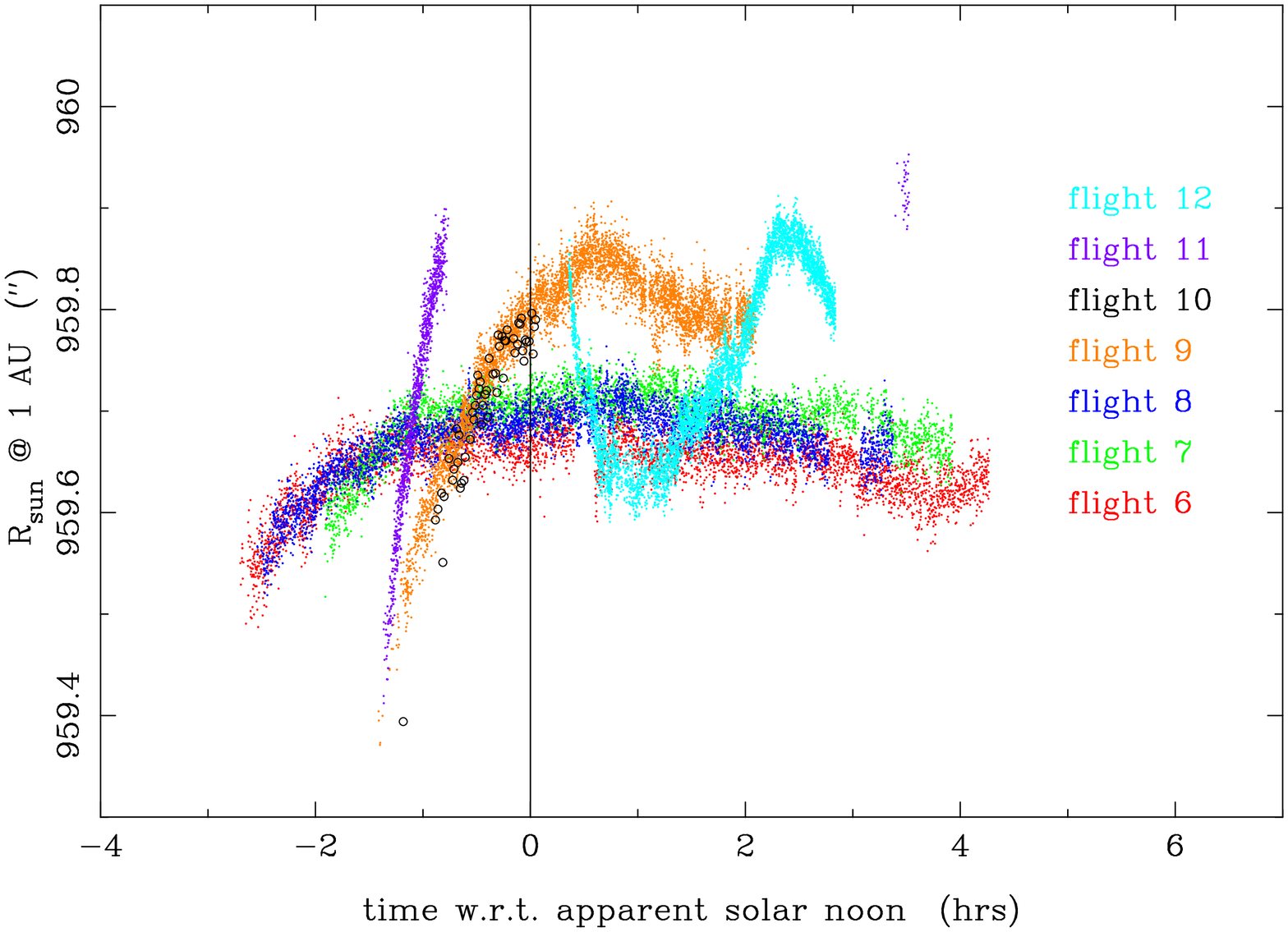}}
\caption{
Solar half-diameter (corrected to 1 AU) as a function of time, relative to
apparent noon, for the seven SDS flights presented here.
The flights are color-coded as indicated in the legend.
Only those portions of each flight that qualify for the determination of
$R_{sun}$ are shown, i.e., direct-image FWHM $<5''$ and $-2''<\Delta$FWHM $<2''$.
For the sake clarity, only one in every 25 such qualifying exposures is plotted,
with the exception of Flight 10 for which all such exposures are shown.
}
\label{F-rsun}
\end{figure}

\begin{figure}
\centerline{\includegraphics[height=0.8\textwidth,clip=,angle=-90]
{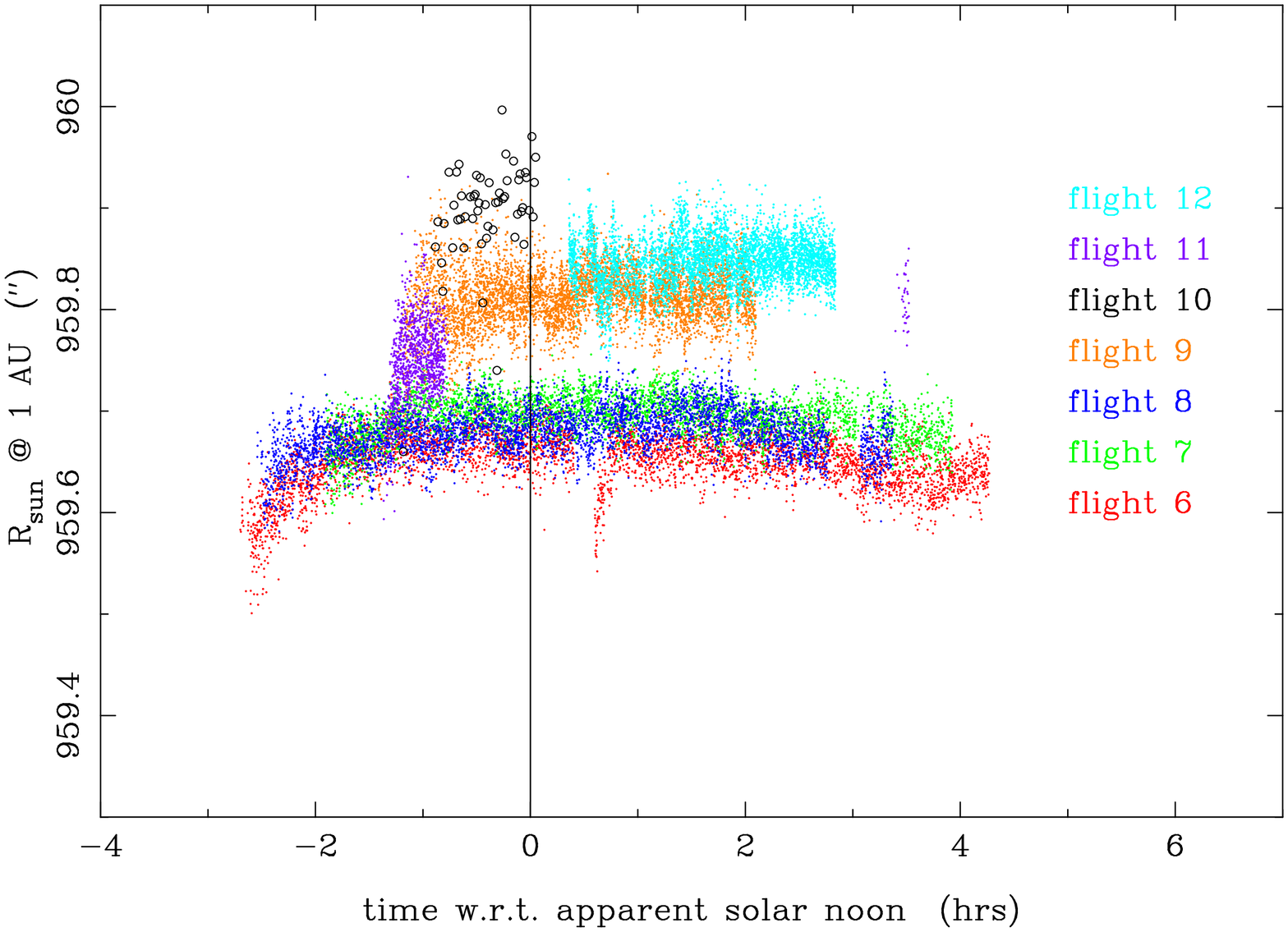}}
\caption{
Same as Figure~\ref{F-rsun} but {\it after} having corrected for the
dependence on limb-width difference, $\Delta$FWHM.
}
\label{F-rsun-del}
\end{figure}

Correcting for the trends seen in Figure~\ref{F-delwid} with a simple linear or
quadratic fit (the results are largely insensitive to the form), the final
measures of $R_{sun}$ throughout the seven flights are derived.
Figure~\ref{F-rsun} shows the in-flight variation $R_{sun}$ prior to making the
final $\Delta$FWHM correction.
Figure~\ref{F-rsun-del} shows the post-correction values.

\section{Solar-radius results}
\label{S-R}

The final SDS measures of the solar half-diameter, after making all
necessary corrections and adjustments described in the previous section,
are listed in Table~~\ref{T-results}.
Dates of observation are expressed in fractional years and the
half-diameter measures are given in arcsec.
The uncertainties listed are $1\sigma$ estimates of the combined random and
time-dependent systematic errors.\footnote[1]{
There is a constant systematic error that we shall ignore, that
of the ``as-built'' reference angle of the BSW.
We are interested in time variation of the solar diameter and, thus, it is
systematic effects that might vary with time that are of concern.
Whatever its exact value, the wedge angle is assumed constant.
}
In all cases but one, (that one being Flight 10), the systematic component is
dominant.
We estimate the systematic uncertainties from the variation of
$R_{sun}$ values resulting from limiting cases of assumptions related to
the corrections listed in Section~\ref{S-DA}, primarily the treatment of
optical distortion.
That is, reductions were made in which the assumed optical distortion was
as designed and with perfect alignment, as well as with distortion fields
assumed to be offcentre and/or with a Barlow-lens misalignment chosen
to force the ratio of direct- and reflected-image radii to unity.
The standard deviation of the $R_{sun}$
value differences was 20 mas and we adopt
this as the systematic component of the uncertainties.
In the case of Flight 10, the small number of recoverable exposures taken
at float altitude resulted in a formal random uncertainty of up to 20 mas,
depending on specifics of the data trimming.
We have chosen, conservatively, to add the random and systematic error
contributions for Flight 11 directly, instead of in quadrature, resulting
in a total estimated uncertainty of 40 mas for this measure.
The majority of the data from Flight 11 also had to be discarded because of
the anomalous behavior of the limb-width during much of the flight, but
even with these data removed, the formal random error was
insignificant relative to the estimated 20-mas systematic component.
To be specific, the formal {\it random} error for Flight 10 is 20 mas; for
Flight 11, it is 0.3 mas; and for the remaining flights it is less than
0.1 mas.
The {\it systematic} component of the total uncertainties listed in
Table~\ref{T-results} is assumed to be the same for each flight and is
estimated to be 20 mas.

\begin{table}
\caption{
SDS Solar-radius results}
\label{T-results}
\begin{tabular}{rrc}
\hline
Flt. & Epoch      &    $R_{sun} @$ 1 AU $('')$     \\
\hline
  6   &  1992.82   &  959.638 $\pm$ 0.020  \\
  7   &  1994.81   &  959.675 $\pm$ 0.020  \\
  8   &  1995.82   &  959.681 $\pm$ 0.020  \\
  9   &  1996.85   &  959.818 $\pm$ 0.020  \\
 10   &  2001.83   &  959.882 $\pm$ 0.040  \\
 11   &  2009.87   &  959.750 $\pm$ 0.020  \\
 12   &  2011.86   &  959.856 $\pm$ 0.020  \\
\hline
\end{tabular}
\end{table}

Note that the present results supersede all previous analyses and
presentations of SDS solar-diameter measures.
Those previous studies included only a portion of the components of the
present analysis; most importantly, they lacked the final adjustment based
on the difference between direct- and reflected-image limb widths.
Another difference of the current analysis is that besides addressing
the known optical processes that affect the results, we have explicitly
considered the uncertainties corresponding to our treatment such processes.

\section{Discussion}
\label{S-D}

\begin{figure}
\centerline{\includegraphics[height=0.8\textwidth,clip=]{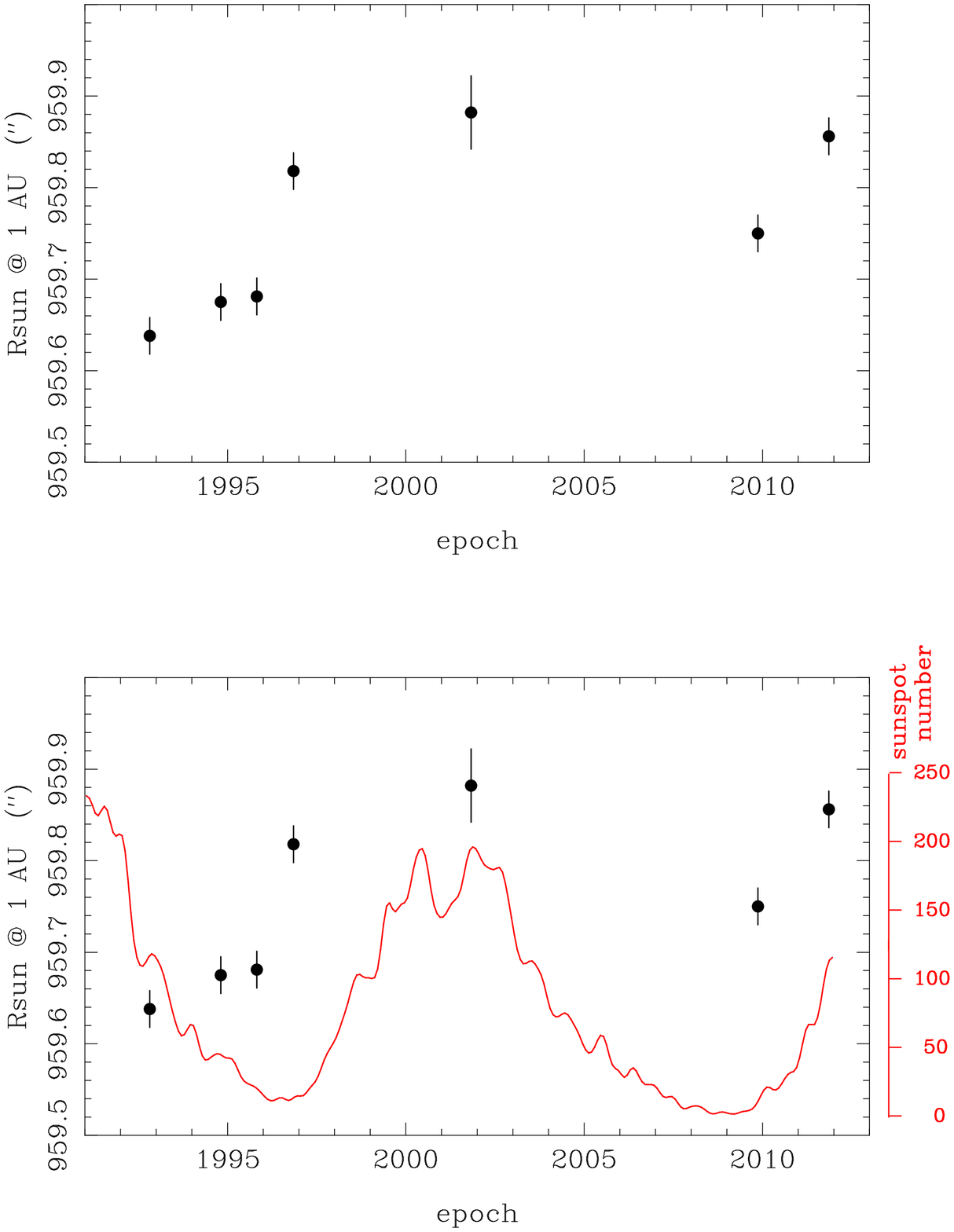}}
\caption{
Final values of the SDS-measured solar half-diameter (corrected to 1 AU)
as a function of time.
The error bars indicate $1\sigma$ estimates of the combined random and
systematic uncertainties.
For comparison, in the lower panel, the SDS results are plotted along with
NOAA-tabulated monthly sunspot numbers for the same period.
}
\label{F-results}
\end{figure}

The $R_{sun}$ values from Table~\ref{T-results}
are plotted in the top panel of Figure~\ref{F-results}.
Based on the estimated precision and stability of the SDS, a significant
variation of the photospheric solar radius with time is apparent.
Are there alternative explanations for the measured variation,
apart from an intrinsic change in the Sun's size? 

The most obvious explanation is that the SDS internal calibration is failing, 
in some way, from flight to flight.
As we have emphasized, the stability of the wedge -- critically, that of the angle
between the second and third surfaces responsible for the reflected image --
is the basis of our astrometric calibration.
SDS construction uses fused silica and Zerodur optical components and structural
supports made of Invar to minimize thermally induced flexure during its Sun-pointed
observing runs.
Still, as indicated by the variations in the limb-edge widths shown in 
Figure~\ref{F-allwid}, the SDS optical properties are seen to vary throughout each 
flight, particularly so during the last four flights.
To the extent that the direct and reflected image paths vary in unison, the 
width-bias correction and pixel-scale adjustment of Section~\ref{S-DA} compensate 
for the expected effect on the limb edge positions, provided the limb-width 
excursions are not overwhelming.
For this reason we discard frames in which the FWHM of the direct image on the
central detector exceeds 5 arcsec.

More troubling are the portions of flights in which the direct and reflected image
properties vary substantially from one another, the bulk of Flight 11 being the
most egregious example of this behavior, (see Figure~\ref{F-allwid}).
A temperature-sensitive ray tracing model of the SDS was explored in an attempt to 
understand how the
reflected-image limb width might maintain nominal behavior while the direct-image
limb width performed such a wild excursion. 
(The inverse behavior would be more readily understood, as a perturbation associated
with the two extra reflections experienced by the reflected beam.)
We found that the mid-Flight-11 widths could be modeled by a specific combination of 
thermal gradients in the first element of the BSW;
a 1$-^{\circ}$C radial gradient (edge to center), and 
a 5$-^{\circ}$C axial gradient (front to back surface).
The former induces a negative power in the first BSW element, affecting both the
direct and reflected images.
This is a result of both a physical deformation of the fused silica element
(coefficient of thermal expansion, $\alpha_{TE} = 5.2 \times 10^{-7} /^{\circ}$C) 
and of the temperature dependence of its index of refraction 
($dn/dT = 1.0 \times 10^{-5} /^{\circ}$C).
The axial gradient, via thermal expansion, also induces a shape change of the
reflecting back surface of the first BSW element, which, if ``tuned,'' can cancel
the refractive effect of the radial gradient and bring the reflected beam back
into focus while the direct beam remains out of focus.
Yet, this explanation also predicts that the effective focal lengths of the direct
and reflected channels would differ by 11\% while in this state.
Such a large scale difference would be readily seen in the ratio of derived disc
radii for the direct and reflected images and nothing anywhere near this level is
observed.
Also, while these balancing thermal gradients could account for the mid-flight widths,
it is highly improbable that the onset and eventual dissipation of the gradients,
as evidenced by the normal width behavior at the start and end of the flight, 
would maintain such a tricky balance, i.e., the reflected-image width behavior never 
deviates far from that seen in other flights.
Thus, we feel this model does not, specifically, explain the extreme behavior during 
the middle of Flight 11.
However, qualitatively, thermally induced shape and/or refractive perturbations 
of the BSW from its equilibrium configuration
are a likely explanation for the observed differences in the limb-width
behavior of the direct and reflected images throughout each flight.
If so, one would expect a deviation from the BSW's equilibrium shape to produce a
deviation in the resulting solar diameter measure.
For small perturbations, the correlation should be linear, exactly as we find
empirically in the limb-width difference adjustment of Section~\ref{S-DAn} 
and Figure~\ref{F-delwid}.
The critical issue, as far using the BSW as a calibration for SDS, is if its
equilibrium shape varies from flight to flight.

The high-altitude environment and thermal load associated with pointing at the
Sun should be relatively consistent for the SDS, from flight to flight.
On the other hand, one can expect a slow deterioration of the reflective coatings
on the BSW surfaces with time, thus, possibly affecting the instrument's in-flight 
equilibrium state under similar thermal stress.
The deterioration of
a subset of the surface coatings can be monitored by examining the relative
heights of the reflected and direct solar images, $\Gamma$, from Equation 2.
Using the same portions of each flight that were used to determine the mean solar
diameter estimates, i.e., filtering by direct-image width and difference in
reflected and direct image widths on the central detector, we list the mean relative
heights in Table~\ref{T-rcap}.
The rms variation of $\Gamma$ within each flight is typically $\sim$0.01.
Overall, there is a gradual trend with time; a slightly larger fraction of light makes
it into the reflected image.
To see if this results in a detectable flight-to-flight change in the equilibrium 
shape of the wedge, we examine the offset of the ghost image limb edge from that of 
the main image, (see Figure~\ref{F-fourpts} and the discussion in 
Section~\ref{S-DAd}).
Recall that the ghost image is due to an extra internal reflection within an
element of the BSW;
the front and back surfaces were purposefully made non-parallel to avoid fringing.
Again restricting ourselves to the useful portion of each flight, the mean values
of $\Delta_{ghost}$ are included in Table~\ref{T-rcap}.
The values are in arcseconds, having been adjusted using simultaneous measures
of the pixel scale based on the separation of the solar disc centres.
The rms within each flight is roughly 0.2 arcsec.
To the extent that a radial thermal gradient changes the shape of the front
BSW element, the ghost image offset should vary by a proportional amount.
In fact, the mean ghost offset is remarkably stable over the final five flights.
What variation is observed, does not correlate with the variation in the measured
solar half-diameter values.
We repeat the mean half-diameter estimates in the table for comparison.

\begin{table}
\caption{
BSW diagnostics by flight}
\label{T-rcap}
\begin{tabular}{rrccc}
\hline
 Flt. & Epoch  & $\Gamma$  & $\Delta_{ghost} ('')$ & $R_{sun} ('')$     \\
\hline
  6   &  1992.82 & 0.661 & 22.80 &  959.638  \\
  7   &  1994.81 & 0.665 & 23.15 &  959.675  \\
  8   &  1995.82 & 0.676 & 22.97 &  959.681  \\
  9   &  1996.85 & 0.698 & 22.97 &  959.818  \\
 10   &  2001.83 & 0.692 & 23.02 &  959.882  \\
 11   &  2009.87 & 0.672 & 22.97 &  959.750  \\
 12   &  2011.86 & 0.720 & 22.97 &  959.856  \\
\hline
\end{tabular}
\end{table}

From this, we find no direct evidence that the BSW geometry, and
thus the SDS calibration angle, is changing from flight to flight.
On the other hand, the underlying cause of the variation in the direct- and
reflected-image limb widths throughout each flight remains not fully understood.
A comprehensive structural/thermal/optical analysis might shed further light, but
at this point resources for an accurate modelling of the as-built SDS are not
available.
Such an model could not be relied upon to make quantitative, ``open-loop'' 
corrections to the instrument scale and distortion, for instance, but it might 
provide a theoretical basis for the empirical adjustments that we employ in our 
reduction procedure.

In the above analysis, $\Gamma$ is sensitive to changes in the reflection/transmission
properties of the coatings on the second and third BSW surfaces while $\gamma$ is
associated with internal reflections from potentially both BSW elements.
All four surface coatings combine to define the passband, roughly 100 nm wide and
centred around 600 nm. 
It is conceivable that with a deterioration of the surface coatings the 
passband could change.
Judging from its roughly constant signal level, the throughput                     
of the instrument has not changed
drastically over the seven flights, although this has not been measured precisely.
Considering the relatively wide passband of the SDS, the effective depth of the
Sun's atmosphere being observed will not be strongly wavelength dependent.
Thus, it is unlikely that evolution of the instrument's passband is responsible for
the observed half-diameter changes.
Besides, one would expect deterioration to yield a monotonic change in the passband
while the observed half diameter is seen to fluctuate.

Even if the BSW-based calibration of the SDS is reliable, there is another possible 
explanation for the apparent variation in the solar diameter worth addressing.
The variation might be caused by
measurement error associated with solar surface activity, should a surface feature
at the Sun's limb fall on one or more of the SDS detectors.
Of course, the frequency and density of such features will fluctuate throughout 
the solar cycle, possibly giving rise to a false diameter change signature.
This explanation is disproved by
the fact that the diameter variations are not in-phase with the activity cycle,
as shown by the lower panel in Figure~\ref{F-results}.  Here the measured
half-diameter values are plotted alongside a tracing
of the NOAA-tabulated sunspot counts over the same period.

Additionally, the SDS is stepped through rotations during each flight,
such that the diameter measures are made at different solar latitudes.
Measurements at all latitudes are included here, thus lessening any
possible dependence on low-latitude activity.
(It is in the determination
of oblateness, the subject of a future study, that an effect due to activity
is likely to be seen, if at all.)
By averaging the SDS measures over large fractions of each flight, the
mean radius over a range of solar latitude is actually what is determined.
For most of the SDS flights
the effective mean solar colatitude measured, $\overline{\theta_l}$, is 
extremely consistent.
The exceptions to this are Flight 10 (loss of most data due to a recorder 
failure), Flight 11 (rejection of anomalous limb profile data), and
Flight 12 (restricted rotation sampling due to reduced telescope control).
The expected flight-to-flight variation in measured radius due to this effect, 
assuming an elliptical limb shape, is $\delta cos(2\overline{\theta_l})$, where 
$\delta$ is the actual difference in equatorial and polar radius.
In Flights 6 through 9 the geometric factor, $cos(2\overline{\theta_l})$, 
is less than 0.1.
Assuming a nominal solar oblateness of 8 mas would induce a systematic
variation from flight to flight that is negligible.
Even in Flights 10 through 12, the factor is at most 0.4, leading to a maximum
systematic offset of 3 mas from the uneven sampling in solar latitude.
This is still insignificant relative to the estimated 20 mas systematic 
uncertainty of each flight's radius determination.
No discernible variation in measured radius as a function
of SDS telescope rotation angle (or in solar latitude observed) is seen
beyond the expected $\sim$10 mas variation associated with true oblateness.
Thus, we conclude that neither solar activity or solar oblateness  -- nor, more
directly, instrument observing angle -- can
account for the flight-to-flight diameter variation observed.

In summary, we find it reasonable to conclude that what the SDS has detected are 
real changes (both increases and decreases) of the solar radius.  These are 
too small to be confirmed by ground-based telescopes, and too few to ascertain 
its time behavior.  Our results contrast with the only alternative space-borne 
radius values published to date (Kuhn et al.~2004), 
which indicate no radius changes 
over nearly two decades.  However, they were obtained with the MDI experiment 
on SOHO, which does not have on-board calibration.

We note that f modes of the Sun can also be used to find a measure of the solar
radius.
Frequencies of solar f modes bear a very simple relationship to stellar
structure.
They are most sensitive to the total mass and radius of the Sun and not to the
details of solar structure.
Solar models constructed with the conventional value of the solar radius
usually have f-mode frequencies that do not match the Sun.
This gave rise to the concept of the ``seismic'' radius of the Sun
(Schou et al.~1997, Antia, 1998).
The seismic radius basically defines where the kinetic energy of the modes is
the maximum.
The seismic radius of the Sun appears to be somewhat smaller than the
photospheric radius.
More interestingly, helioseismic data show that the Sun's seismic radius
varies with changes in solar activity
(Antia et al. 2000);
in the sense
that the seismic radius decreases with increase in solar activity.
By comparison,
the photospheric radius variations (i.e., those presented here, as
determined from limb position)
are much larger and not in phase with the activity cycle.
This can be understood by the fact that the radius variations are not
homologous
(Sofia et al. 2005; Lefebvre et al.~2007),
with different mass
shells expanding by different amounts.

Simultaneous measurements of the photospheric and seismic radii are of great
interest since they will allow us to calibrate solar models better.
Differences and similarities of the temporal behavior of these radii should
contain information on how and where magnetic fields linked to activity
change solar structure (Sofia \& Li 2001, Li et al.~2003).

\section{Conclusions}
\label{S-C}

Accurate, long-term measurement of the diameter of the Sun is an
admittedly difficult undertaking, fraught with issues of instrument
stability and calibration.
The SDS design overcomes these challenges by relying on a beam splitting
wedge, constructed using optical contact bonding, to provide side-by-side
images of the Sun separated by a stable reference angle.
The dual images transform the relatively large angular measure into a
small displacement in the instrument's focal plane, and also provide
a number of diagnostic parameters that can be used to monitor and internally
calibrate the diameter measurements throughout each SDS flight.
The SDS data-analysis procedure, presented here in detail, allows direct
comparison of results over the decades-long program.

The balloon-borne SDS experiment has measured the angular size of the
Sun seven times over the period 1992 to 2011.
The half-diameter is found to change over that time by up to 200 mas, whereas
the estimated uncertainties of the measures, random plus systematic, are
typically 20 mas.
The variation is {\it not} in
phase with the solar activity.
Thus, the measured variation is not
an artefact of observational contamination by surface activity.
While the SDS measures span 19 years, they are sparse, making it impossible
to say with any certainty that the observed variation is cyclic.

The temporal behavior of the Sun's photospheric radius provides a key
constraint on models of solar structure, particularly with regard to the
location, geometry, and evolution of subsurface magnetic fields.
Because radius variations might have significant implications regarding the 
effects of solar variability on climate change (Sofia \& Li 2001), it is necessary 
to make further efforts to confirm or refute the changes reported in this 
paper, to further refine its magnitude, and to establish its time behavior.  
This requires a long series of observations which are well calibrated and made 
from space, or a space-like environment.

%

An experiment that has spanned nearly three decades is only possible with
the support and collaboration of many people and institutions.
Beyond the
co-authors of this paper, we would first like to thank the people who assisted
in its conception and design;
this includes E.~Maier, K.~Schatten, P.~Minott,
H-Y.~Chiu, and A.~Endal.
The fabrication of the SDS payload could not have been
accomplished without D.~Silbert.
Flight operations were helped by many people
including D.~Pesnell, and W.~Hoegy.
We are, of course, grateful for the substantial financial
support provided by a series of grants from NASA and the NSF,
and more recently, from the G.~Unger Vetlesen
Foundation, and the Brinson Foundation.
CNES (France) and CNRS (France),
which support the PICARD/SODISM mission,
also provided financial support for the most recent SDS flight and we
thank these institutes for their contribution.
Understanding of the SDS instrument response was greatly assisted by
the COSI model provided by A.~Shapiro and optical modelling by 
L.~Ramos-Izquierdo as well as modelling and other contributions by
W.~van Altena, R.~Mendez and D.~Casetti.
Our work benefited from previous versions of the SDS analysis pipeline
constructed by J.~Zhang, A.~Egidi, B.~Caccin, and D.~Djafer.
Also, insightful comments from an anonymous referee led to significant
improvements in the manuscript.
Finally, we want to acknowledge the
outstanding flight
support provided by the personnel of the NASA/Columbia Scientific Balloon
Facility over the years.


\appendix
\section{Profile skew and $IPP$ shift}

As discussed in Section~\ref{S-DAf}, broadening of the limb edge profile
by a symmetric kernel function results in a shift of the $IPP$ because of
an asymmetry in the underlying, intrinsic limb profile.
This shift is characterized by the measured FWHM as shown in Figure~\ref{F-fitsim}.
Additional $IPP$ shift or bias will be incurred if the broadening mechanism 
itself imparts an asymmetry, i.e., if the instrument point spread function
is asymmetric.

\begin{figure} 
\centerline{\includegraphics[height=0.5\textwidth,clip=]{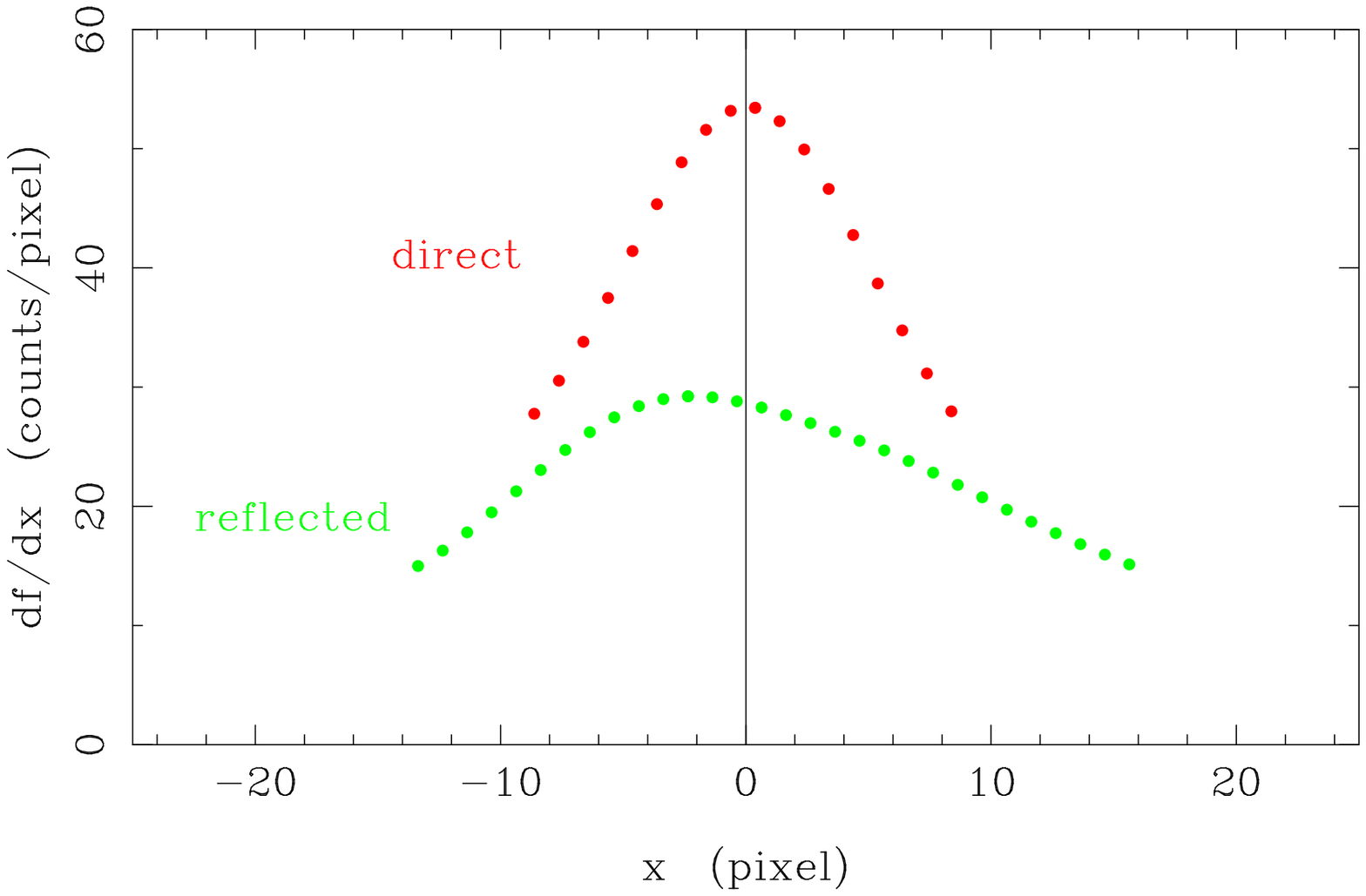}}
\caption{
Sample limb-edge profile pair in which the reflected-image limb shows skew while
the direct-image limb does not.
The numerical derivative of the smoothed profile is from a single exposure
on the central detector, midway through Flight 12.
$x$=0 corresponds to the Gaussian-fit center for each limb, while the FWHM is
2.1 and 4.1 for the direct and reflected images, respectively.
}
\label{F-skew} 
\end{figure}

Such asymmetry, beyond that expected from the intrinsic limb profile, is seen
in the SDS observations when the FWHM rises above its nominal 2 to 3 arcsec value.
Figure~\ref{F-skew} illustrates this for an exposure during Flight~12.
Numerical derivatives of the smoothed limbs, measured on the central detector,
are shown for a case where the direct and reflected images yield FWHM values of
2.1 and 4.1 arcsec, respectively.
There is significant skew in the broader, reflected-image profile.
The local inflection point differs from the Gaussian-fit center ($x$=0 in the
figure) by 2 to 3 pixels, while the expected shift using the calibration curve
of Figure~\ref{F-fitsim} is only 1 pixel.
The instrument (in this case, its reflected-image channel) must be responsible 
for the additional asymmetry.

A better fit to such a profile can be made by employing an asymmetric
fitting function.
Alternatively, a symmetric fitting function can still be used provided the range
of data points to be included in the fit is sufficiently limited.
Recall that the portion of the profile to be fit was 
somewhat arbitrarily chosen to be 
those data points contiguous to the maximum point and with values greater than 
half the maximum.
As a test, we choose the second approach, adopting the quadratic defined by the
maximum point in each profile and its neighboring two points.
While the use of just three points per profile will make for noisy $IPP$ 
determinations, it is systematic effects we wish to explore here.

\begin{figure}
\centerline{\includegraphics[height=0.9\textwidth,clip=]{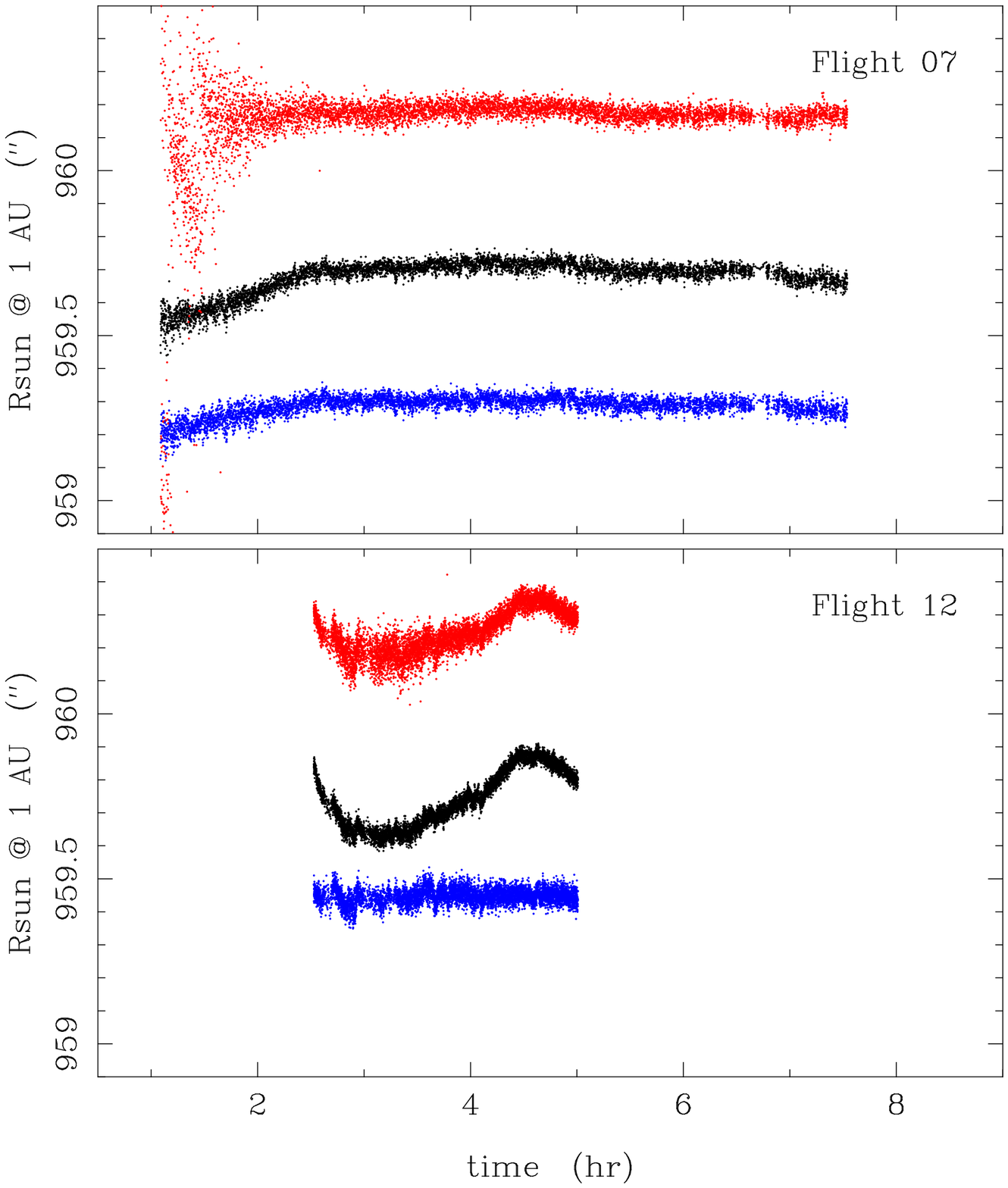}}
\caption{
Comparison of $IPP$-fitting methods for Flight~07 (upper panel) and 
Flight~12 (lower panel).
Solar half diameter, corrected to 1 AU, is shown as a function of the raw in-flight
time, which has an arbitrary zero point.
In each panel, the central (black) points show the $R_{sun}$ measures based on the
standard pipeline but before
the final $\Delta$FWHM correction, while the lower (blue) points show the same
data after the $\Delta$FWHM correction.
Using the alternative, three-point quadratic-fit $IPP$ method results in the
half diameter measures shown by the upper (red) set of points.
The latter sets of points are artificially displaced by +/-0.3 arcsec for the
sake of clarity. 
}
\label{F-alx}
\end{figure}

We apply this $IPP$-determination method to Flights 07 and 12, and plot the
resulting half-diameters in Figure~\ref{F-alx}, (red points).
For comparison, we also plot the half diameters resulting from the standard
(Gaussian fit) pipeline, before (black points) and after (blue points) the final
$\Delta$FWHM correction of Section~\ref{S-DAn}.
An artificial offset of +/-0.3 arcec has been added to separate the three sets for
clarity.

Assuming the Sun's true angular diameter is constant throughout each flight, it
appears the three-point quadratic $IPP$ method does an equally good job
of flattening the flight curve as does the $\Delta$FWHM-correction method.
In fact, in terms of systematics, if not random noise, it is superior during the
initial ascent phase of the flight.
Although, during this stage the FWHM value exceeds 5 arcsec and, thus, the
standard pipeline would have rejected these data for the purpose of diameter
determination.

For Flight 12, the conclusion is different; the three-point quadratic $IPP$ 
method does not yield a constant half diameter while the standard pipeline,
with the $\Delta$FWHM correction, does.
Note that during Flight 07 the direct and reflected limb widths track one
another well, while during Flight 12 they do not, (see Figure~\ref{F-delwid}).

\end{document}